\documentclass{article}
\usepackage[onecolumn]{emulateapj}
\usepackage{flushrt}
\usepackage{psfig}

\makeatletter
\@ifundefined{chapter}{\def\thebibliography#1{\section*{References\@mkboth
  {REFERENCES}{REFERENCES}}\list
  {\relax}{\setlength{\labelsep}{0em}
        \setlength{\itemindent}{-\bibhang}
        \setlength{\itemsep}{0pt}
        \setlength{\parsep}{0pt}
        \setlength{\leftmargin}{\bibhang}}
    \def\newblock{\hskip .11em plus .33em minus .07em}
    \sloppy\clubpenalty4000\widowpenalty4000
    \sfcode`\.=1000\relax}}%
{\def\thebibliography#1{\chapter*{Bibliography\@mkboth
  {BIBLIOGRAPHY}{BIBLIOGRAPHY}}\list
  {\relax}{\setlength{\labelsep}{0em}
        \setlength{\itemindent}{-\bibhang}
        \setlength{\itemsep}{0pt}
        \setlength{\parsep}{0pt}
        \setlength{\leftmargin}{\bibhang}}
    \def\newblock{\hskip .11em plus .33em minus .07em}
    \sloppy\clubpenalty4000\widowpenalty4000
    \sfcode`\.=1000\relax}}

\newlength{\bibhang}
\let\@internalcite\cite
\def\cite{\let\@citeleft(\let\@citeright)%
    \@ifstar{\citeyear}{\citefull}}
\def\citenp{\let\@citeleft\relax\let\@citeright\relax
    \@ifstar{\citeyear}{\citefull}}
\def\citefull{\def\astroncite##1##2{##1~##2}\@internalcite}
\def\citeyear{\def\astroncite##1##2{##2}\@internalcite}

\def\@citex[#1]#2{\if@filesw\immediate\write\@auxout{\string\citation{#2}}\fi
  \def\@citea{}\@cite{\@for\@citeb:=#2\do
    {\@citea\def\@citea{; }\@ifundefined
       {b@\@citeb}{{\bf ?}\@warning
       {Citation `\@citeb' on page \thepage \space undefined}}%
{\csname b@\@citeb\endcsname}}}{#1}}

\def\@cite#1#2{\@citeleft#1\if@tempswa , #2\fi\@citeright}
\def\@biblabel#1{}

\makeatother

\newcommand\approxgt{\, \mbox{$^{>}\hspace{-0.24cm}_{\sim}$} \,}
\newcommand\approxlt{\, \mbox{$^{<}\hspace{-0.24cm}_{\sim}$} \,}

\begin{document}

\title{On Estimating the QSO Transmission Power Spectrum}  
\author{Lam Hui\altaffilmark{1}, Scott Burles\altaffilmark{2},
Uro\v s Seljak\altaffilmark{3}, Robert E. Rutledge\altaffilmark{4},\\
Eugene Magnier\altaffilmark{5} and David Tytler\altaffilmark{6}}
\altaffiltext{1}{Institute for Advanced Study, Olden Lane, Princeton,
NJ 08540; NASA/Fermilab Astrophysics Center, Fermi
National Accelerator Laboratory, Batavia, IL 60510; e-mail: \it
lhui@ias.edu}
\altaffiltext{2}{Department of Astronomy and Astrophysics, University
of Chicago, Chicago, IL 60637}
\altaffiltext{3}{Physics Department, Princeton University, Princeton,
NJ 08544} 
\altaffiltext{4}{Department of Astronomy, California Institute of
Technology, Pasedena, CA 91125}
\altaffiltext{5}{Canada-France-Hawaii Telescope,
65-1238, Mamalahoa Hwy,
Kamuela, HI  96743}
\altaffiltext{6}{Center for Astrophysics and Space Sciences, University
of California, San Diego, MS 0424, La Jolla, CA 92093-0424
}

\keywords{methods: data analysis --- intergalactic medium --- quasars:
absorption lines --- cosmology: observations --- large scale structure
of universe}

\begin{abstract}
The Lyman-alpha forest has become an important tool for measuring
the mass power spectrum at high redshifts ($z = 2-4$).
A crucial intermediate step is the measurement of the 
transmission power spectrum.
We present new methods to minimize the systematic and random errors
for such a measurement,
and discuss their implications for
observing strategies. Sources of systematic errors explored include 
metal line contamination and continuum-fitting. We advocate
the technique of trend-removal in place of 
traditional continuum-fitting -- here, a spectrum is normalized
by its (smoothly varying) mean rather than its continuum
-- this method is easily automated and removes biases
introduced by continuum-fitting. Moreover, trend-removal
can be easily applied to spectra where continuum-fitting
is difficult, such as when the resolution or signal-to-noise
is low, or for spectra at high redshifts. 
We further show that a measurement
of the continuum power spectrum (plus a related quantity) using
trend-removal,
from either low redshift quasar spectra or the red-side of Lyman-alpha,
can be used to constrain the amount of spurious large scale
power introduced by the uncertain continuum, and in principle
allows the removal of such contamination and thereby expanding
scales probed to larger ones.
We also derive expressions for the shot-noise bias and
variance of the power spectrum estimate, taking
into account the non-Poissonian nature of the shot-noise
and the non-Gaussianity of the cosmic fluctuations.
An appropriate minimum-variance weighting of the data is given. 
Finally, we give practical suggestions
on observing strategy: the desired resolution and S/N
for different purposes and instruments, and how to distribute 
one's finite observing time among quasar targets. Also discussed
is the quasar spectroscopic study of the Sloan Digital Sky Survey 
(SDSS), which has the potential to measure the power spectrum
accurate to better than $1 \%$ per mode ($\Delta k \sim 10^{-4} {\, 
\rm s/km}$) -- the techniques presented here will be useful for 
tackling the anticipated issues of shot-noise and continuum 
contamination.
\end{abstract}

\section{Introduction}
\label{intro}

Recent theoretical research on the low column density ($N_{\rm HI}
\approxlt 10^{14.5} {\rm cm^{-2}}$) Lyman-alpha (Ly$\alpha$) forest
at redshifts $z \sim 2 - 4$ points towards a picture in which
the forest consists largely of
mildly nonlinear fluctuations of a smooth intergalactic medium
(e.g. \citenp{bbc92,cen94,zhang95,rm95,hernquist96,jordi96,muecket96,bi97,bond97,croft97,hui97a,hg97};
see \citenp{rauch98} for a review and further ref.).
This provides the motivation to analyze the quasar (QSO) absorption 
spectrum
as a continuous field with fluctuations, rather than as a collection of
discrete absorption lines. The two-point correlation or its fourier
transform, the power spectrum, comes to mind as a useful and common
statistic used in other areas such as the microwave background or
galaxy large-scale-structure. Indeed, its application to QSO
spectrum has been discussed by a number of authors
(\citenp{zuo94,jordi96,bi97,cen98}). Recently, Croft et al.
\cite*{croft98,croft98b} (see also \citenp{hui99,mcdonald99}) 
have shown that the
mass power spectrum can be recovered from the QSO transmission power
spectrum, from which one could further deduce cosmological parameters
such as $\Omega_m$ (\citenp{weinberg98,phillips00}). 
There exist at present a large number of high quality QSO spectra 
(e.g. \citenp{hu95,lu96,kirkman97,cristiani97,kim97,rauch97}) which
makes this an exciting field of research. Upcoming quasar surveys
such as the Sloan Digital Sky Survey (SDSS) and the Anglo-Australian
Telescope Two Degree Field (AAT2DF) will enlarge the database significantly.

Here we take the view that the QSO transmission power
spectrum / correlation is interesting in its own right, and focus on
how to best measure it from the observed QSO spectra, independent of
the underlying physical picture of the forest.
The two major questions are: (1) what are the main sources of systematic
errors and what are the best ways to bring them under control? (2) how
to estimate the shot-noise, and 
to best combine data with different signal-to-noise (random errors)?

Let us start by defining the transmission power spectrum and
correlation function. Two possibilities arise. One of them
we call the un-normalized power spectrum $P_{\rm un}$ /
two-point correlation
$\xi_{\rm un}$ (Weinberg 1998, private communication; 
\citenp{mcdonald99}):
\begin{equation}
\xi_{\rm un} (u) = \langle f (u') f (u' + u) \rangle \, \, , \,
\, P_{\rm un} (k) = \int \xi_{\rm un} (u) e^{-iku} du
\label{Pxidef0}
\end{equation}
where $f$ is the transmission defined by $f = e^{-\tau}$ with $\tau$ being the
optical depth (the absorption is then $1-f$), $u$ or
$u'$ is the observed velocity (or redshift or wavelength) along a
line of sight, and $k$ is its
fourier counterpart. The angular brackets denote ensemble averaging.

The other we call the normalized power spectrum $P$ / two-point correlation
$\xi$ (e.g. \citenp{zuo94}):
\begin{equation}
\xi (u) = \langle \delta_f (u') \delta_f (u' + u) \rangle \, \, , \,
\, P (k) = \int \xi (u) e^{-iku} du
\label{Pxidef}
\end{equation}
where $\delta_f = (f - \bar f) / \bar f$, with 
$\bar f$ being the mean transmission $\langle f \rangle$.

We will almost exclusively focus on the latter, but will
discuss at some point the pros and cons of the two, especially with
regard to systematic errors. Unless otherwise
stated, hereafter, power spectrum / correlation refers to the
normalized version. 

The layout of the paper is as follows.
In \S \ref{overview}, we provide a brief overview of how the raw data
output (a two-dimensional CCD image) is reduced to a one-dimensional
QSO spectrum. Note that the quantity $f$ above is never observed directly. It is
important to have a description of how the whole data reduction
procedure works, which is sometimes hard to find in the literature.
We give some illustrations by showing simulated spectra with various realistic levels of noise. 

In \S \ref{estimating}, we discuss the estimation of the power
spectrum and two-point correlation, beginning with the introduction
of the quadratic estimator in \S \ref{estimator}. 
An important point pertaining to the estimation of the two-point
correlation is raised here -- most estimators employed in the
literature are not optimal; an alternative is given here which
is an analogue of the Landy-Szalay estimator (\citenp{ls93})
introduced originally for galaxy surveys.
Aside from this point, we focus exclusively on the estimation of
the power spectrum. In \S \ref{systematics}, we discuss 
three sources of systematic errors:
continuum-fitting, gaps and metal absorption lines. 
Particular attention is paid to issues related to continuum-fitting.
We advocate in \S \ref{continuum} trend-removal to replace traditional
continuum-fitting, which avoids the latter's pitfalls.
We further propose in \S \ref{beyond} that the power spectrum of
the continuum can be estimated using trend-removal as well, 
which offers us a way to measure accurately the transmission
power spectrum on large scales where the continuum fluctuations
might be important. In \S \ref{gaps}, we discuss the effects
of gaps and un-removed metal lines.
We then turn our attention in \S \ref{random} to random errors.
We emphasize here that the shot noise is not exactly Poisson 
distributed because of the particular way the data is reduced.
We point out the importance of subtracting the shot-noise-bias
correctly, and 
describe a systematic way of assigning error-bars to the power
spectrum, and introduce minimum-variance-weighting techniques to
combine data of different qualities. Some results here are
stated without justification. The aim in this section is to
summarize useful results for readers who might not be interested 
in details of the derivations, which are provided in the Appendices.
The techniques used in the Appendices should be of broad interest
e.g. the issue of generally non-Poissonian shot-noise might be 
relevant for galaxy power spectrum estimation.

Lastly, we conclude in \S \ref{conclude}. We summarize
here our recipe for transmission power estimation -- 
readers who would like a quick overview of our methods can
skip directly to this section, and only refer back to the
relevant sections to fill in the details. We give general
advice on observing strategies, and 
discuss in particular analysis issues relevant for
the Sloan Digital Sky Survey. 

Before we begin, let us first make a
few clarifying remarks about some of our notation and terminology.

\section{Terminology and Notation: Averaging and Averages}
\label{notation}

In this paper, we refer to two different kinds of fluctuations which
should be clearly distinguished. Take the observed photon (or
electron) count from a quasar
as an example, $\hat N_{\rm Q}$. As one moves along a spectrum,
the photon count fluctuates because of two very 
different reasons. First, it fluctuates because the universe
is intrinsically inhomogeneous, giving rise to non-uniform absorption
-- we will refer to these as cosmic fluctuations. Second, it
fluctuates because the observed photon count is a discrete realization
of the underlying cosmic signal -- Poisson fluctuation is the
canonical example but not the only possible one, we will refer
to these as discrete fluctuations. 

We define two different kind of averages corresponding to these
two different kind of fluctuations. The discrete average of the
observed photon count is denoted by $\langle \hat N_{\rm Q} \rangle_D
\equiv \tilde N_{\rm Q}$. In other words, $\hat N_{\rm Q}$ 
constitutes a discrete realization of the underlying cosmic signal $\tilde
N_{\rm Q}$. This signal $\tilde N_{\rm Q}$ itself suffers from cosmic
fluctuations, and we will denote its ensemble average by 
$\langle \tilde N_{\rm Q} \rangle = \bar N_{\rm Q}$. 
A fixed quasar continuum is assumed in this ensemble average i.e. it
is the fluctuation in the spectrum caused by intervening
absorption that constitutes the
cosmic signal we are after. Finally, we will sometimes use
$\langle \, \rangle$ to implicitly
stand for $\langle \langle \, \rangle_D \rangle$ e.g.
$\langle \hat N_{\rm Q} \rangle$ actually means
$\langle \langle \hat N_{\rm Q} \rangle_D \rangle$ which is the same
as $\langle \tilde N_{\rm Q} \rangle = \bar N_{\rm Q}$
\footnote{Two exceptions in the use of $\langle \, \rangle$:
in \S \ref{beyond}, we use $\langle \, \rangle$ to include
averaging over an ensemble of different continua, and
in \S \ref{random} we use $\langle \, \rangle_{kk}$ to denote
averaging over a shell of Fourier modes.}

To recap:
\begin{itemize}
\item $\hat N_{\rm Q}$ is the directly observed quasar photon count.
\item $\tilde N_{\rm Q} \equiv \langle \hat N_{\rm Q} \rangle_D$ refers
to the idealized quasar photon count if one has infinite signal-to-noise
(S/N).
\item $\bar N_{\rm Q} \equiv \langle \tilde N_{\rm Q} \rangle
= \langle \langle \hat N_{\rm Q} \rangle_D \rangle$ refers
to the quasar photon count if one has infinite S/N {\it and} if
one averages over all possible
cosmic fluctuations keeping the continuum fixed e.g. by taking
the same quasar and putting it at all possible orientations in the sky.
For instance, if $\tilde N_{\rm Q} = N_C e^{-\tau}$ where $N_C$
is the true continuum and $\tau$ is the optical depth, then
$\bar N_{\rm Q} = N_C \langle e^{-\tau} \rangle$ where 
$\langle e^{-\tau} \rangle$ is the mean transmission.
\end{itemize}

Note that, when we use the term discrete, it is not implied that
$\hat N_{\rm Q}$ is necessarily an integer, although it
is derived from some integral quantity such as the 
electron/photon count.
We use the term discrete fluctuations  instead of the usual Poisson
fluctuations, because as we will see, $\hat N_{\rm Q}$ is often not
Poisson-distributed i.e. 
$\langle \hat N_{\rm Q}^2 \rangle_D - \langle
\hat N_{\rm Q} \rangle_D^2 \ne \tilde N_{\rm Q} = \langle 
\hat N_{\rm Q} \rangle_D$
(see \S \ref{describe}). The term shot-noise is often
used to describe Poissionian discrete fluctuations, but in this
paper we will use it more broadly to include non-Poissonian
discrete fluctuations as well.

Finally, we note that we use the term 'quasar counts' throughout
this paper to refer to the photon counts from a quasar, rather
than the number of quasars in a given patch of sky.

\section{Data Reduction: From the CCD Image to the QSO Spectrum}
\label{overview}

\subsection{A Brief Description}
\label{describe}

We discuss briefly here aspects of the data-processing
necessary for understanding the noise
properties of the reduced quasar spectrum. The reader is referred
to Horne \cite*{horne86}, Zuo \& Bond \cite*{zuo94}, Barlow \&
Sargent \cite*{barlow97}, Rauch et al.
\cite*{rauch97} and Cen et al. \cite*{cen98} for 
more discussions. 

The raw data consist of a two-dimensional (spatial and spectral) array
of counts (data numbers or photon counts converted from them) from a
CCD image. The one-dimensional array of estimated quasar counts in the
spectral direction (as a function of velocity, redshift or wavelength)
is obtained by 
collapsing the data in the spatial direction in the following fashion:
\begin{equation}
\hat N_{\rm Q}^\alpha = \sum_{i,\beta}W^{\alpha\beta} W^{i\beta} (\hat
N_{\rm RAW}^{i\beta} 
- {\tilde N}_{\rm 
B}^{i\beta})
\label{Ncount}
\end{equation}
We have introduced and will stick with the following notations:
the Latin index such as $i$ and the Greek index such as $\alpha$
denote the spatial and spectral coordinates respectively of a CCD
pixel (there are in fact a few exceptions, which should
be clear from the context); 
$\hat
N_{\rm Q}^\alpha$ is our
estimated quasar count, $\hat N_{\rm
RAW}^{i\beta}$ is the raw count, and ${{\tilde N}_{\rm B}}^{i\beta}$ is
the mean background count which includes the sky and the readout
offset; $W^{i\beta}$ is a weighting of the spatial pixels for each spectral
coordinate $\beta$, and $W^{\alpha\beta}$ represents a rebinning
of the spectral pixels that is sometimes done to achieve, for instance,
a linear wavelength scale. Note that the $\alpha$ and $i$
dimensions do not necessarily 
align with the two perpendicular axes of the CCD chip.
The optical setup could be such that the spectrograph slit
appears tilted at an angle to the CCD axes.
We use $\hat{}$ to emphasize the fact that the quantity of interest is a
random variable with fluctuations. The $\tilde{}$ denotes a discrete
average: e.g. ${{\tilde N}_{\rm B}}^{i\beta} = \langle {{\hat N}_{\rm
B}}^{i\beta} \rangle_D$ where $\langle \, \, \rangle_D$ denotes discrete
averaging. 

Implicitly assumed in the above formulation is that 
the discrete average ${{\tilde N}_{\rm B}}^{i\beta}$ is known, which is
of course not strictly true, but since a typical slit covers
a significant number of pixels that do not have any quasar photons,
and since the background is often quite uniform, the discrete
average can be estimated to high accuracy. Note also the readout
offset can be measured
using short exposures with closed
shutters or from the CCD overscan region.

The weighting $W^{i\beta}$ typically has non-trivial spectral
dependence ($\beta$) to remove at least two artifacts:
variations in detector efficiency across the chip and 
a non-flat blaze. The former is usually estimated in a procedure
called flat-fielding by shining a lamp into the detection system.
The latter arises because of the non-trivial shape of a 
diffraction order. This can be partially estimated in the flat-fielding
procedure, but is best done using a spectrophotometric standard star,
usually a white dwarf.
While the correction for the first artifact should be quite accurate,
the blaze-removal is often approximate.
Any residual that is not correctly removed
will show up in the form of a non-trivial effective continuum.
We will see in \S \ref{systematics} perhaps some evidence of it.
We assume in this paper that such artifacts show up as 
fluctuations on large scales (since the blaze itself is smoothly and
slowly varying across a given order) but not on small scales (we will
quantify the scales later on). 

To make the above concrete, the raw count is given by:
\begin{equation}
\hat N_{\rm RAW}^{i\beta} = \hat N_{\rm B}^{i\beta} + \hat N_{\rm Q}^{i\beta}
\label{raw}
\end{equation}
where the quasar contribution has the following discrete average:
\begin{equation}
{\tilde N}_{\rm Q}^{i\beta} \equiv \langle \hat N_{\rm Q}^{i\beta}
\rangle_D = g_{\rm ps}^{i\beta} g_{\rm 
b}^{\beta} \tilde N_{\rm Q}^{\beta} 
\label{ggNQ}
\end{equation}
where $g_{\rm ps}^{i\beta}$ is the point-spread function which describes how
the light from the quasar gets spread-out in the spatial direction $i$ at
a given spectral coordinate $\beta$, and $g_{\rm b}^\beta$ accounts
for the variation of the blaze and quantum efficiency
as a function of wavelength. The symbol ${\tilde N}_{\rm Q}^{\beta}$
denotes the underlying quasar count (or cosmic signal i.e. discrete averaged). 

Many different rebinning kernels $W^{\alpha\beta}$ (eq.
[\ref{Ncount}]) are possible. 
%A reasonable  choice would be for
%the kernel to satisfy $\sum_{\beta}
%W^{\alpha\beta} = 1$. 
The simplest choice is of course no
rebinning with $W_{\alpha\beta} = \delta_{\alpha\beta}$. 

There are several possible choices of the weighting $W^{i\beta}$ (eq.
[\ref{Ncount}]), but any sensible choice has to satisfy the
requirement that
$\langle \hat N_{\rm Q}^\alpha \rangle_D = \tilde N_{\rm Q}^\alpha$,
up to some constant normalization factor.
This assumes that artificial fluctuations
introduced by the blaze or detector efficiency are correctly taken
out. If not, it shows up effectively as part of the continuum.

We give two examples of $W^{i\beta}$ here. 
The first is basically a uniform weighting over the spatial pixels
that correspond to a given spectral coordinate:
\begin{equation}
W^{i\beta} = 1/(g_{\rm b}^{\beta} \sum_j g_{\rm
ps}^{j \beta}) \, ,
\label{W1}
\end{equation}
where the range of $i$, or $j$, is chosen to lie within, say, some
fraction of the quasar seeing disk. There is sometimes an additional 
complication due to cosmic ray hits, which will 
be discussed below.

The second is a minimum variance weighting (different from minimum 
variance
weighting for measuring the power spectrum; \S \ref{random}) over
the spatial pixels, introduced by Horne \cite*{horne86}:
\begin{equation}
W^{i\beta} = (1/g_{\rm b}^{\beta}) (g_{\rm
ps}^{i \beta}/ V_{\rm 
RAW}^{i \beta}) / (\sum_j [g_{\rm ps}^{j \beta}]^2 / V_{\rm
RAW}^{j \beta})
\label{W2}
\end{equation}
where $V_{\rm RAW}^{j \beta}$ is the variance in the
raw count:
\begin{eqnarray}
\label{vRAW}
&& V_{\rm RAW}^{j \beta} = \langle (\hat N_{\rm RAW}^{j \beta})^2 \rangle_D -
\langle \hat N_{\rm RAW}^{j \beta} \rangle_D^2 = \tilde N_{\rm Q}^{j \beta}
+ V_{\rm B}^{j \beta} \\ \nonumber
&& {V_{\rm B}}^{j \beta} = {\tilde N_{\rm S}}^{j\beta} + V_{\rm R.O.}^{j\beta}
\end{eqnarray}
where $V_{\rm B}^{j \beta}$, the background variance, has two
contributions, the sky variance ${\tilde N_{\rm S}}^{j\beta}$
and the readout variance $V_{\rm R.O.}^{j\beta}$.
A word of caution is necessary here regarding the second weighting. 
The raw variance, $V_{\rm 
RAW}^{j\beta}$, depends on the underlying cosmic signal or quasar
count
($\tilde N_{\rm Q}^{j\beta}$ i.e. discrete averaged) 
which is not directly observable (the discrete averaged sky count
and the true readout variance are also strictly speaking not directly
observable, but they can be estimated quite accurately because 
they are relatively uniform and can be observed over a larger number of
pixels). Modeling $V_{\rm RAW}^{j\beta}$ using the measured raw
count (i.e. using $\hat N_{\rm Q}^{j\beta}$ instead of $\tilde N_{\rm
Q}^{j\beta}$ in eq. [\ref{vRAW}]) could lead to a biased estimation of
${\tilde N_{\rm 
Q}^\beta}$. Horne \cite*{horne86} suggested an iterative scheme to
avoid this problem, 
but implementations of this weighting should be checked for a possible bias.

Pixels affected by cosmic-ray hits, which are usually easy to
identify because of
their wild fluctuations and spiky nature, are dealt with in two
different ways, depending on the severity. For a given spectral
coordinate, if only a small fraction of
the corresponding spatial pixels are affected, the weighting in eq.
(\ref{W1}) or eq. 
(\ref{W2}) is simply modified by allowing $i$ and $j$ to only range
over the unaffected spatial pixels.
However, if all or most corresponding spatial pixels are affected,
then all recorded counts at that spectral coordinate are discarded,
leaving a gap in the reduced quasar spectrum. Gaps could result also
because of metal-line removal (an alternative would be to fit for
the metal-line and subtract, instead of simply discarding the pixels)
or defects in the CCD.

Finally, the (random) error-array output at the end of the data reduction
corresponds to an estimate of
\begin{equation}	
\sqrt {\langle (\hat N_{\rm Q}^\alpha - {\tilde N_{\rm Q}}^\alpha)^2
\rangle_D} =   
\sqrt {\sum_{i,\beta} (W^{\alpha\beta} W^{i{\beta}})^2
V_{\rm RAW}^{i\beta}} 
\label{deltapN}
\end{equation}	
where we have assumed that the noise-fluctuations are independent
among the pixels. We emphasize that in practice the error-array is only 
an {\it
estimate} of the above quantity, because the true $V_{\rm
RAW}^{i\beta}$ is unknown, but is estimated using
the observed raw counts (using $\hat N_{\rm Q}^{j\beta}$ instead of
$\tilde N_{\rm Q}^{j\beta}$ in eq. [\ref{vRAW}]).

It is clear from the above discussion that in general
fluctuations in $\hat N_{\rm Q}^\alpha$ are non-Poissonian,
in the sense that $\langle \hat N_{\rm Q}^2 \rangle_D - \langle
\hat N_{\rm Q} \rangle_D^2 \ne \langle \hat N_{\rm Q} \rangle_D 
= \tilde  N_{\rm Q}$. This is because of several reasons.
First, $\hat N_{\rm Q}$ suffers from additional discrete fluctuations
from the background counts. Second, the weighting $W^{\alpha\beta}$ and
$W^{i\beta}$ are in general non-trivial (i.e. not unity). A
very simple example of the effect of non-unit weights is: suppose we
multiply a Poisson variable by a factor of 2 and call
the result $\hat y$, it is easy to see that $\langle \hat y^2  \rangle_D
- \langle \hat y \rangle_D^2 = 2 \langle \hat y \rangle_D \ne
\langle \hat y \rangle_D$.

For the rest of this paper, we will pick for simplicity the
weighting kernels $W_{\alpha\beta} = \delta_{\alpha\beta}$  and $W_{i\beta}$ as given
by eq. (\ref{W1}). In the Appendices
we will indicate where some of our expressions have to be modified
to account for more general weightings.

\subsection{Simulated QSO Spectra}
\label{simulations}

\begin{figure}[htb]
\centerline{\psfig{figure=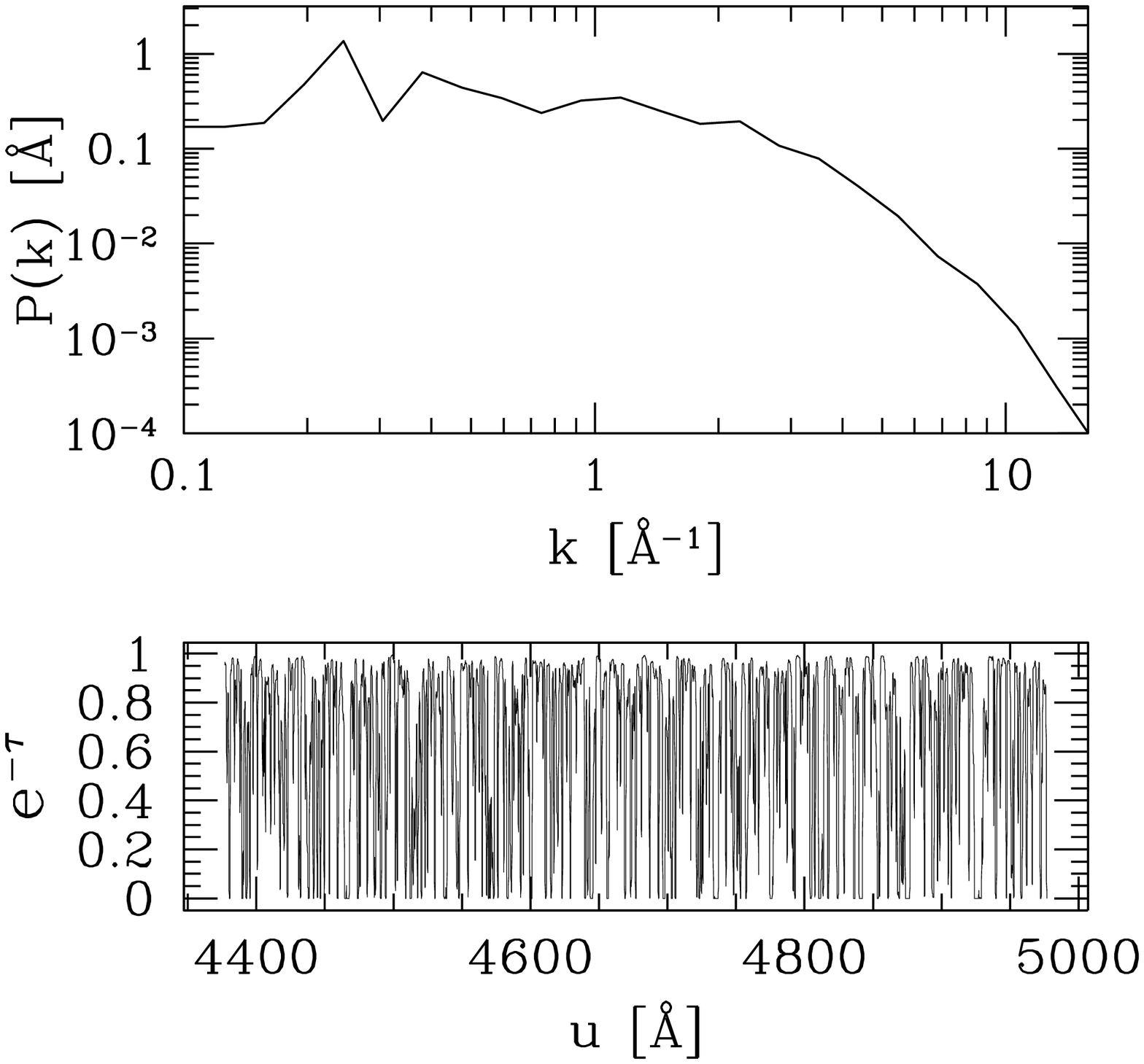,height=5.0in}}
\caption{The lower panel shows the transmission $e^{-\tau}$ as a
function of wavelength $u$ taken from a SCDM simulation, with no
noise added. The upper panel shows the corresponding (normalized) transmission 
power spectrum (eq. [\ref{Pxidef}]). All subsequent simulated spectra
in this paper are based on this one, with various levels of noise,
contamination, etc added.} 
\label{NOnoise}
\end{figure}

For illustrations, and for later analyses, we have generated several
different QSO spectra. The underlying noiseless (theoretical)
transmission ($f = e^{-\tau}$) is shown in the bottom panel of Fig.
\ref{NOnoise}, and its associated power spectrum is shown in
the top panel of the same figure.
This is drawn from a SCDM (Standard Cold Dark Matter) simulation 
discussed in Gnedin \cite*{gnedin98} which made use of the Hydro-PM
algorithm developed 
by Gnedin \& Hui \cite*{gh98}. The cell-size (comoving size of $10 {\,
\rm h^{-1} 
kpc}$) is small enough to resolve the effective
Jeans scale, and so should retain all small scale structures.
However, the box-size is unrealistically small (comoving size $2.56
{\, \rm h^{-1} Mpc}$) which means a significant amount of large scale
power is missing.  
For most of our investigations here, it is not necessary that the 
simulations
are highly realistic, but our simulated transmission power
spectrum is in fact broadly consistent with an observed one
(Fig. \ref{demonstrate}).
The long line of sight in Fig. \ref{NOnoise} is generated
by shooting a ray at some oblique angle through the simulation
box and allowing it to wrap around the box several times, but never repeating
itself. The mean redshift here is $z = 2.85$. The ionizing background
is chosen to give $\langle e^{-\tau} \rangle = 0.64$ (\citenp{press93}).

%NOnoise.plot
%\begin{figure}[htb]
%\centerline{\psfig{figure=NOnoise.ps,height=7.0in}}
%\caption{The lower panel shows the transmission $e^{-\tau}$ as a
%function of wavelength $u$ taken from a SCDM simulation, with no
%noise added. The upper panel shows the corresponding (normalized) transmission 
%power spectrum (eq. [\ref{Pxidef}]). All subsequent simulated spectra
%in this paper are based on this one, with various levels of noise,
%contamination, etc added.} 
%\label{NOnoise}
%\end{figure}
%%shows simulated spectrum from 
%%/d/sculptor/lhui_sculptor/PDS/NewErr/MoreData/Long0.037_NO.noise.gap.co%%nt
%%the power spectrum is from
%%/d/honus/lhui/Run/MoreRun/Long0.037_NO.noise.gap.cont/N1_Nkk30_Ispds0tr0
%%seems O.K.

An example of a somewhat realistic reduced QSO spectrum ($\hat N_{\rm
Q}^\alpha$ in eq. [\ref{Ncount}]) and its error
array (eq. [\ref{deltapN}]) can be found in the bottom two panels of
Fig. \ref{highKeckNM}
(ignore the other two panels for the moment).
They are generated based on the prescriptions given in \S
\ref{describe}, assuming $W^{\alpha\beta} = \delta^{\alpha\beta}$ and
$W^{i\beta}$ is given by eq. (\ref{W1}). Briefly speaking, what we do
is to first generate an array of $g^\alpha$ which
represents $g_{\rm b}^\alpha \sum_j g_{\rm ps}^{j\alpha}$ (i.e. we do not
actually simulate the full two-dimensional CCD image; the spatial
dimension is collapsed into $g^\alpha$); then, we create a Poisson
realization of the (intermediate) quasar count $g^\alpha N_C^\alpha
e^{-\tau_\alpha}$ where 
$N_C$ is the continuum and $e^{-\tau_\alpha}$ is predicted by our
cosmological model; we similarly create a Poisson realization of
the background count $g^\alpha \times {\rm const.}$ where the constant
represents some fraction less than 1 and then subtract from it its
Poisson mean, the end-result is then added to the above quasar count;
lastly, we divide by $g^\alpha$ to obtain the reduced quasar count $\hat
N_{\rm Q}^\alpha$.

Note that the overall level of the reduced
quasar count can be 
scaled up or down (because we are not interested in the absolute
brightness of the quasar), provided the error array is scaled accordingly to
conserve signal-to-noise (S/N; count divided by square root of the
variance).
This example resembles a high quality Keck HIRES spectrum, with
S/N reaching up to 100 at certain pixels. 
It is composed of 12 echelle orders, $50 \AA$ each (e.g. the instrument
HIRES on the Keck telescope is an echelle
spectrograph 
which consists of two diffraction gratings crossed at $90^0$ to each
other; see Vogt et al. \citenp*{vogt94}). The pixel size is $0.05 \AA$ with a
resolution Full-Width-Half-Maximum (FWHM)
of $0.125 \AA$. The example represents a case in which
a relative calibration (but not necessarily absolute fluxing) between
the orders has been attempted. The dashed line in the bottom panel
shows the input continuum. 

The error-array shows a lot of variations. 
A model of the random error as Gaussian distributed with uniform
S/N that is sometimes used in the literature misses much structure.
About $3 \%$ of the spectrum consists of gaps which arise
due to severe cosmic-ray hits.
The spikes in the error-array correspond to wavelengths at which
a fraction, but not all, of the corresponding spatial pixels are
affected by cosmic-ray hits. They also take up $3 \%$ of the spectrum.
It is easy to see how these spikes arise from eq. (\ref{W1}) and
(\ref{deltapN}). At wavelengths where some of the spatial pixels are
thrown out because of cosmic-ray hits, $W_{i\beta}$ is enhanced
because the sum over $j$ in its denominator is restricted to fewer
pixels. Since it is the square of $W_{i\beta}$ that
enters into the variance, a modest enhancement becomes a spike.
Note how for each echelle order, the
S/N drops towards the two ends. This is because of the
blaze function which tends to suppress the flux at the ends.
Note also that the S/N has a general decline towards the
blue. This is due to a combination of a falling continuum, and
decreasing detector efficiency.

%highKeckNM.C.plot
%\begin{figure}[htb]
%\centerline{\psfig{figure=highKeckNM.C.ps,height=7.0in}}
%\caption{Similar to Fig. \ref{highKeckNM} except that a relative
%calibration between different echelle orders have not been done, and
%the input continuum is taken from continuum-fits to an observed quasar
%spectrum. The dotted line in the top two panels corresponds
%to the case where the continuum is modeled as flat for each order.
%The dashed line is where polynomials up to the third order are used
%to fit for the mean transmission in each order. The solid line in the
%second panel from the top shows the true mean transmission.} 
%\label{highKeckNM.C}
%\end{figure}
%spectrum from 
%/d/sculptor/lhui_sculptor/PDS/NewErr/MoreData/Long0.037_simerr6_cr_Nf1_cont3NM
%continuum is from cont3.dat
%
%estimated mean's from
%/d/lyra/lhui/Run/MoreRun/Long0.037_simerr6_cr_Nf1_cont3NM
% /N1_Nkk20New_Ispds1Istr1
%/d/lyra/lhui/Run/MoreRun/Long0.037_simerr6_cr_Nf1_cont3NM
% /N4_Nkk20New_Ispds1Istr1
%
% DP/P computed using as true power spectrum
%/d/lyra/lhui/Run/MoreRun/Long0.037_NO.noise.gap.cont
% /N1_Nkk20New_Ispds1Istrend1, instead of /N1_Nkk20New_Ispds1.

Sometimes, a relative calibration between echelle orders is
either difficult or simply not attempted. An example is shown in 
Fig. \ref{highKeckNM.C}. Note how the continuum is broken into
12 discontinuous pieces.\footnote{In some cases where
the different echelle orders overlap, there could be two jumps
at each order junction.} 
These are taken from continuum-fits to actual
data.

%LickNM_cont3L.plot
%\begin{figure}[htb]
%\centerline{\psfig{figure=LickNM_cont3L.ps,height=7.0in}}
%\caption{The bottom 2 panels are the same as in Fig.
%\ref{LickNM_cont3L_S}. The second panel from the top shows the
%recovery
%of the mean transmission assuming a model composing of
%polynomials up to the third order (dashed line). The solid line
%shows the true mean. The dashed line in the top panel
%represents $(\hat P_{2} - P)/P$ -- the error for the normalized power
%spectrum estimated using eq. (\ref{hatP}). The dotted line
%shows the power spectrum estimate if shot-noise were not subtracted.}
%\label{LickNM_cont3L}
%\end{figure}

%see pdsScottCont3.plot (the old version used pdsScottCont.plot)
%and /d/lyra/lhui/LA-compute/Analyze/Test3-31-99/guide, look for
%Runs/Cont

An example of data with much poorer quality is shown in Fig.
\ref{LickNM_cont3L}. The pixel size is $0.5 \AA$ and the FWHM is $1.17
\AA$. The S/N is about 10 times worse than the two
examples above. Such a spectrum could be the output of, say, a low
dispersion single-grating spectrograph, which does not
have the characteristic division into short pieces as in the case
of the echelle spectrograph.

All other simulated data in this paper are slight variations of the above,
which will be described in turn at the appropriate places.

\section{Estimating the Power Spectrum / Two-point Correlation}
\label{estimating}

\subsection{The Quadratic Estimator}
\label{estimator}

Given the one-dimensional array of estimated quasar counts $\hat
N_{\rm Q}^\alpha$ (eq. [\ref{Ncount}]), how should one go about
estimating the two-point correlation or the power spectrum?

A common practice is to first continuum-fit, i.e. to estimate the
continuum level $\hat N_{\rm C}^\alpha$ 
and divide $\hat N_{\rm Q}^\alpha$ by it to obtain an estimate
of the transmission $\hat f = \hat N_{\rm Q}^\alpha / \hat N_{\rm
C}^\alpha$. Then, the estimators for the un-normalized
two-point correlation and power spectrum (eq. [\ref{Pxidef0}]) are:
\begin{equation}
\hat \xi_{\rm un} (u) = \sum_{\alpha,\beta} w^{\alpha\beta} (u)
\hat f^\alpha \hat f^\beta \, \, , \, \, 
\hat P_{\rm un} (k) = \sum_{\alpha,\beta} w^{\alpha\beta} (k)
\hat f^\alpha \hat f^\beta - b (k) 
\label{hatP0}
\end{equation}
where $b (k)$ subtracts out the shot-noise (i.e. a bias), and
$w^{\alpha\beta} (u)$ and $w^{\alpha\beta} (k)$ are weighting
kernels for which we will give some examples shortly (to be
distinguished from $W^{\alpha\beta}$ in eq. [\ref{Ncount}]).
These are commonly called quadratic estimators for the simple fact
that they make use of the data $\hat f^\alpha$ in
quadratic combinations. 

Defining the mean transmission to be $\bar f$,
the obvious extensions of the above estimators, for the normalized
two-point correlation and power spectrum (eq. [\ref{Pxidef}]; unless
otherwise stated, the 
two-point correlation or power spectrum with no qualifications refer
to the normalized version), are
\begin{equation}
\hat \xi_1 (u) = \sum_{\alpha,\beta} w^{\alpha\beta} (u) (\hat f^\alpha
-{\bar f}) (\hat f^\beta
-{\bar f}) / {\bar f}^2 \, \, , \, \,
\hat P_1 (k) = \sum_{\alpha,\beta} w^{\alpha\beta} (k) (\hat f^\alpha
-{\bar f}) (\hat f^\beta
-{\bar f}) / {\bar f}^2
- b (k)
\label{hatPOLD}
\end{equation}

However, the form of the power spectrum or two-point correlation
estimator given above suggests an interesting variation which allows
us to avoid continuum-fitting altogether: 
$(\hat f^\alpha - {\bar f}) / {\bar f}$ can be estimated instead by
$(\hat N^\alpha - {\bar N^\alpha}) / {\bar N^\alpha}$ where
${\bar N^\alpha}$ is the mean count defined by
$\bar N^\alpha \equiv \langle \hat N^\alpha \rangle$. Here $\langle \,
\rangle$ denotes the cosmic average i.e. this 
corresponds to averaging out the cosmic fluctuations in $f$, for
a fixed continuum ($\langle \hat N^\alpha \rangle = N_C^\alpha \bar f$
where $N_C^\alpha$ is the true continuum). \footnote{It is implicitly
assumed that discrete averaging has been carried out before
cosmic averaging i.e. we use $\langle \, \rangle$ 
interchangeably with $\langle \langle \, \rangle_D \rangle$. See
\S \ref{notation}.}
Note that the mean count is dependent upon $\alpha$ because
of the slowly varying continuum. 
We will discuss how to estimate $\bar N^\alpha$ 
shortly. The key here is that the absolute
level of the continuum gets divided out by definition.
Hence, let us define the following alternative estimators of the
two-point correlation  
and power spectrum: 
\begin{mathletters}
\label{hatP}
\begin{eqnarray}
\label{hatPP}
&& \hat \xi_{2} (u) = \sum_{\alpha,\beta} w^{\alpha\beta} (u)
\hat\delta_f^\alpha \hat \delta_f^\beta \, \, , \, \,
\hat P_{2} (k) = \sum_{\alpha,\beta} w^{\alpha\beta} (k)
\hat\delta_f^\alpha \hat \delta_f^\beta - b (k) \\
\label{hatdeltaf}
&& \hat \delta_f^\alpha \equiv (\hat N_{\rm Q}^\alpha 
-\bar N_{\rm Q}^\alpha) / {\bar N}_{\rm Q}^\alpha
\end{eqnarray}
\end{mathletters}
This alternative power spectrum estimator is what we will focus on,
but we will also briefly investigate the behavior of the estimators in
eq. (\ref{hatP0}) and (\ref{hatPOLD}). 

It remains to specify what $w^{\alpha\beta} (u)$, $w^{\alpha\beta} (k)$
and $b (k)$ are. 
The simplest choice is to use uniform weighting i.e. for the
two-point correlation, it corresponds to:
\begin{eqnarray}
w^{\alpha\beta} (u) = \Theta^{\alpha\beta} (u)/ \sum_{\mu\nu}
\Theta^{\mu\nu} (u)
\label{wu}
\end{eqnarray}
where $\Theta^{\alpha\beta} (u)$ is equal to one if the two pixels
$\alpha$ and $\beta$ are separated by a distance $u$
(or more generally, the distance falls into a bin that is centered
around $u$ with some finite width), and zero otherwise.
Using the above $w^{\alpha\beta} (u)$ corresponds to simply
counting all pairs separated by a distance $u$, normalized by
the total number of pairs.

With the above weighting, eq. (\ref{hatP}) is analogous
to an estimator of the two-point correlation introduced by Landy \&
Szalay \cite*{ls93} for
galaxy surveys: $ (DD - 2DR + RR) / RR$, if 
one identifies $DD$ with $\sum_{\alpha\beta} 
w^{\alpha\beta} (u)\hat N^\alpha \hat N^\beta$, $DR$ with 
$\sum_{\alpha\beta}
w^{\alpha\beta} (u) \hat N^\alpha \bar N^\beta$ and $RR$ with
$\sum_{\alpha\beta} w^{\alpha\beta} (u)\bar N^\alpha
\bar N^\beta$
and assumes $\bar N^\alpha$ varies very slowly with $\alpha$ on
the scale of interest $u$ (the analogy becomes exact in the limit
of a uniform $\bar N^\alpha$).
As shown by Landy \& Szalay \cite*{ls93} (see also
\citenp{szapudi98,dodelson97}), a common alternative 
estimator $DD/RR - 1$ (equivalent to the estimator used by e.g.
\citenp{zuo94,cen98}) 
is actually less desirable as it gives 
a larger variance compared to $ (DD - 2DR + RR) / RR$.

With this being said, we are going to focus our attention
on the power spectrum from now on, although most of our
treatments below can be applied to the two-point correlation as well.
For the power spectrum, the simplest choice of uniform weighting
corresponds to:
\begin{mathletters}
\label{wkbkuniform}
\begin{eqnarray}
\label{wkuniform}
&& w^{\alpha\beta} (k) = ({\cal L}/{\cal N}^2) R^{\alpha\beta} (k)
\, \, , \, \, R^{\alpha\beta} (k) \equiv (1/n_k) \sum_{k'} e^{-ik'(u^\alpha -
u^\beta)}  \\
\label{bkuniform}
&& b (k) = {{\cal L}\over {\cal N}^2} \sum_\alpha
{q^\alpha \bar N_Q^\alpha + 
V_B^\alpha \over (\bar N_Q^\alpha)^2}  \, \, , \, \,
q^\alpha \equiv \sum_i (W^{i\alpha})^2 g_{\rm ps}^{i\alpha} g_{\rm
b}^\alpha \, \, , \, \, V_B^\alpha \equiv \sum_i (W^{i\alpha})^2
V_B^{i\alpha} 
\end{eqnarray}
\end{mathletters}
where ${\cal L}$ is the total length of the spectrum 
(in whatever units one
prefers) and ${\cal N}$ is the total number of spectral pixels,
$R^{\alpha\beta} (k)$ represents an average of the Fourier basis
over some bin or band in k-space i.e. we estimate the power spectrum at a bin
centered at $k$ by averaging over contributions from each $k'$ that belongs
to the bin ($n_k$ is the total number of modes in it).
\footnote{See Seljak \cite*{seljak97} and Bond et al. \cite*{bjk97}
for discussions on precautions one should take on binning.} The symbol
$b (k)$ represents the shot-noise contribution to the power that
has to be subtracted off, $g_{\rm ps}^{i\alpha}$ and $g_{\rm
b}^\alpha$ represent the point-spread function and the blaze as in eq.
(\ref{ggNQ}), and $V_B^\alpha$ is the background contribution to the shot-noise
(including the sky and readout, see eq. [\ref{vRAW}]). 
Note how a part of the shot-noise depends on the reciprocal of the
mean quasar count, as expected for Poisson fluctuations, quite
analogous to the shot-noise of galaxy distributions. \footnote{See e.g.
Feldman et al. \cite*{fkp94}.}
However, the factor $q^\alpha$, which arises from non-trivial
weighting of CCD pixels ($W^{i\alpha}$; eq. [\ref{Ncount}]), signifies
that the shot-noise is not strictly Poisson distributed. Moreover,
there are extra contributions to the shot-noise from the background
counts, which are generally absent in the case of galaxy surveys.
Derivations of the above statements are given in Appendix A.

The corresponding power spectrum estimator obeys:
\begin{eqnarray}
\label{Gwindow}
\langle \hat P_2 (k) \rangle = \int {dk'\over {2\pi}} G(k,k') P(k') \,
\, , \, \,
G(k,k') \equiv \sum_{\alpha\beta} w^{\alpha\beta} (k)
e^{ik'(u^\alpha- u^\beta)}
\end{eqnarray}
where $G$ is a window function that resembles, for 
$k \gg 1/{\cal L}$, a delta function
centered at $k = k'$ with a width of the order of $1/{\cal L}$. 
The normalization of $w^{\alpha\beta} (k)$ in eq. (\ref{wkuniform})
ensures that $\int {dk' G(k,k')/{2\pi}} = 1$. 
See Appendix A for a derivation.

It should be emphasized, however, that the above statements are
strictly true only if one ignores uncertainties in the mean count $\bar
N_Q^\alpha$ -- i.e. $\bar N_Q^\alpha$
is not known a priori but is instead estimated from the
same data from which one is trying to measure correlations.
We will discuss this further in \S \ref{beyond}. It suffices
to say here that our results in this section remain valid
as long as one stays away from scales comparable to the
entire length of the quasar spectrum.

\subsection{Systematic Errors}
\label{systematics}

\subsubsection{Continuum-fitting versus Trend-removal}
\label{continuum}

The power spectrum estimator $\hat P_2$ in eq. (\ref{hatP}), which we
are going to focus most of our attention on, requires an estimate
of mean count $\bar N_Q^\alpha$. The mean count is not strictly uniform
because of a slowly fluctuating continuum i.e. $\bar N_Q^\alpha =
N_C^\alpha \bar f$ where $N_C^\alpha$ is the continuum and $\bar f$ is
the (flat) mean transmission. We assume $\bar N^\alpha$
has the following form: 
\begin{equation}
\bar N_{\rm Q}^\alpha = \sum_a C^a p^{a\alpha}
\label{polyn}
\end{equation}
where $p^0$ is a constant, $p^{1}$ is the first order polynomial
($p^{1\alpha} = u^\alpha$), $p^{2}$ is the second order polynomial
($p^{2\alpha} = (u^\alpha)^2 $), and so on. The coefficients $C^a$ need to
be estimated from the quasar counts $\hat N_Q^\alpha$. 
Note that most of our following arguments would go through for
a different set of basis functions.
One key assumption we will exploit is that $\bar N^\alpha$ is
slowly fluctuating, so that we can truncate the above series at 
relatively low orders. Continuum-fitting in practice makes the same
assumption.

To estimate $C^a$, we use a linear estimator:
\begin{equation}
C^a = \sum_\alpha M^{a\alpha} \hat N_{\rm Q}^\alpha
\label{M}
\end{equation}
where ${\bf M}$ is a matrix to be specified. Comparing eq.
(\ref{polyn}) and (\ref{M}), it is not hard to see that ${\bf M}$ has
to satisfy $\sum_\alpha M^{a\alpha} p^{b\alpha} = \delta^{ab}$. 
The simplest choice is to adopt, in vector notation, ${\bf M = (p
p^T)^{-1} p}$ where ${\bf p p^T}$ in component-form is $[p p^T]^{ab} =
\sum_\alpha p^{a\alpha} p^{b\alpha}$. In summary, this means
our estimator for the mean quasar count is
\begin{equation}
\bar N_{\rm Q}^\alpha = \sum_\beta L^{\alpha\beta} \hat N_{\rm Q}^\beta \, \, , \, \, 
{\bf L} \equiv {\bf p^T (p p^T)^{-1} p}
\label{Nmeanest}
\end{equation}
where ${\bf L}$ in component-form reads
$L^{\alpha\beta} \equiv \sum_{ab} p^{a\alpha} \tilde p^{ab}
p^{b\beta}$ with $\tilde p^{ab}$ being the inverse of the matrix
$[p p^T]^{ab}$. More sophisticated versions of the above can be found
in Rybicki \& Press (\citenp*{rybicki92}). Our experience is that
the simple version given here suffices, because the shape of the true quasar
continuum is quite uncertain anyway. 

Note the crucial differences between traditional continuum-fitting and
an estimation of the mean count 
as described above. The above method makes no reference to the
absolute level of the continuum i.e. the count level where there is
supposedly no absorption. Continuum-fitting in practice often
involves human intervention (eye-balling) in the identification
of such a level.  
In contrast, eq.
(\ref{Nmeanest}) is straightforward to implement and automate.
The mean count is then used to normalize the quasar
count as in eq. (\ref{hatdeltaf}) before the power spectrum is
estimated (eq. [\ref{hatP}]). We call this procedure trend-removal to
distinguish it from traditional continuum-fitting. 
Trend-removal is widely used in other areas (e.g.
\citenp{press92,rybicki92,tegmark98}).
It is akin to the estimation of, say, the long-term trend of the
stock market in the midst of all its daily fluctuations. 

Eq. (\ref{Nmeanest}), together with eq. (\ref{hatP}) and (\ref{wkbkuniform}),
completely specifies the main power spectrum
estimator we advocate. Several tests follow.

{\noindent \it Test 1}

\begin{figure}[htb]
\centerline{\psfig{figure=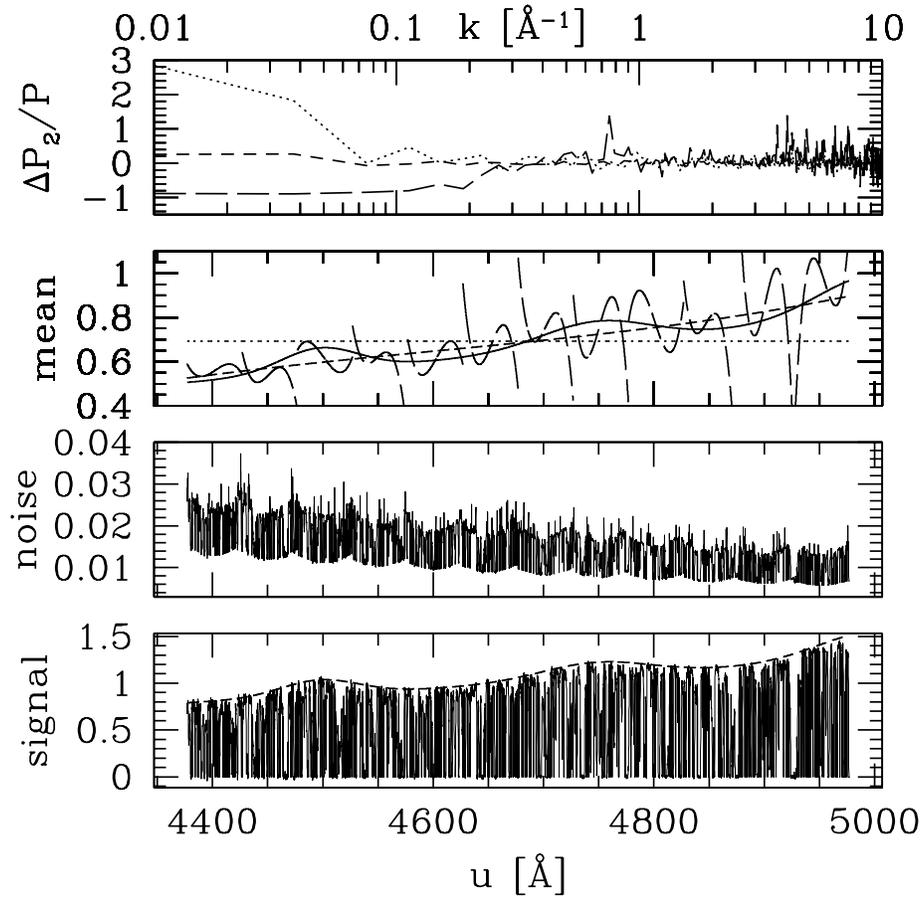,height=5.0in}}
\caption{Bottom two panels show the reduced quasar spectrum (signal)
with its noise array (eq. [\ref{deltapN}]), which resembles
a high resolution (8 km/s FWHM) echelle spectrum with 12
orders, each $50 \AA$ long. It is assumed a relative calibration
between the different orders has been attempted. The spectrum is
generated from the one in Fig. \ref{NOnoise}, with suitable noise
added as described in \S \ref{simulations}. Note that the overall
normalization of the signal or noise is arbitrary but the ratio of the
two is not. The dashed line of the bottom panel shows the input
continuum level. The second panel from the top shows the recovered
mean-transmission for three different choices of the
mean-transmission-basis (${\bf p}$; eq. [\ref{polyn}]): dotted line
for a flat model-continuum/mean ($p^0 = 1$ only); short-dashed line for a
basis of polynomials up to the third order ($p^0$ ... $p^3$);
long-dashed line for a basis that is the same as that for the
short-dashed line except that each echelle order is fitted separately.
The solid line gives the true mean.
The top-panel shows the result of applying the power spectrum
estimator $\hat P_2$ (eq. [\ref{hatPP}]) with uniform weighting
(eq. [\ref{wkbkuniform}]). The symbol $\Delta P_2 /P$ stands for
$(\hat P_2 - P)/P$ where $P$ is the true power spectrum. The different
lines correspond to different choices of the mean-basis, labeled as
in the second panel.}
\label{highKeckNM}
\end{figure}

In Fig. \ref{highKeckNM}, we show the effect of different
choices of the mean-transmission basis ${\bf p}$ (eq. [\ref{polyn}]). 
The simulated spectrum is of Keck-HIRES quality, with a S/N as high as
100 at certain pixels, and it assumes one has a
good relative calibration between the different echelle orders
(12 in all) i.e. an almost ideal, state-of-the-art observed spectrum. 
The second panel from the top shows the recovery of the mean
transmission. The solid line represents the true (input) mean. The rest
shows the recovered mean for different bases: dotted line
for a basis consisting of $p^0$ only (a constant i.e. the continuum or
the mean
is modeled as completely flat); short-dashed line for a
basis consisting of polynomials up to the third order; long dashed
line for also a basis of polynomials up to the third order but
with coefficients fitted separately for each echelle order. 
The short-dashed line seems to give the best match to the true mean.
However, none of them is perfect because the true mean does not, by choice,
have a polynomial shape. This is what is likely to happen
in practice -- lacking a good understanding of the physics
that determines the continuum shape of any given quasar, the best one
can do is to pick a reasonable basis which contains enough freedom to
describe the general features of the continuum, but not so much
freedom that one overfits. 

The important question is what impact
the choice of basis has on power spectrum estimation.
This is illustrated in the top panel of Fig. \ref{highKeckNM}, where
the fractional error in the measured power spectrum is shown.
The one that gives the best match to the true power spectrum
is indeed the one where a simple basis of $p^0$, ..., $p^3$ is used
for the whole length of the simulated spectrum. 
The biggest effect of 
underfitting (dotted
line) or overfitting (long dashed) the mean transmission is on the
power spectrum estimation on large scales. They cause respectively
over- or under-estimation of the large scale power spectrum.
An additional effect is that overfitting tends to introduce spurious
power on small scales as well -- witness the enhanced fluctuations in the error
on small scales for the long-dashed line. We will see this more
clearly in the next Test.

Without any prior knowledge of the intrinsic continuum shape of an
observed quasar, how does one decide if one is overfitting or
underfitting? One way is to look at the region of the observed
spectrum redward of the Ly$\alpha$ emission line, which is free
of the forest, and the continuum is therefore relatively easy to
reconstruct. Assuming the general level of continuum-fluctuation is the same
both redward and blueward of Ly$\alpha$, one can then gain an idea
of what a good mean-transmission basis might be.
Low redshift QSO spectra, where 
the continuum can be quite easily recovered even blueward of
Ly$\alpha$, 
can also be used to 
gauge the scales at which the continuum fluctuates
-- naturally, one could also use such spectra
to check the assumption that continuum-fluctuations have
similar characteristics redward and blueward of Ly$\alpha$ (more on
this below).

{\noindent \it Test 2}

\begin{figure}[htb]
\centerline{\psfig{figure=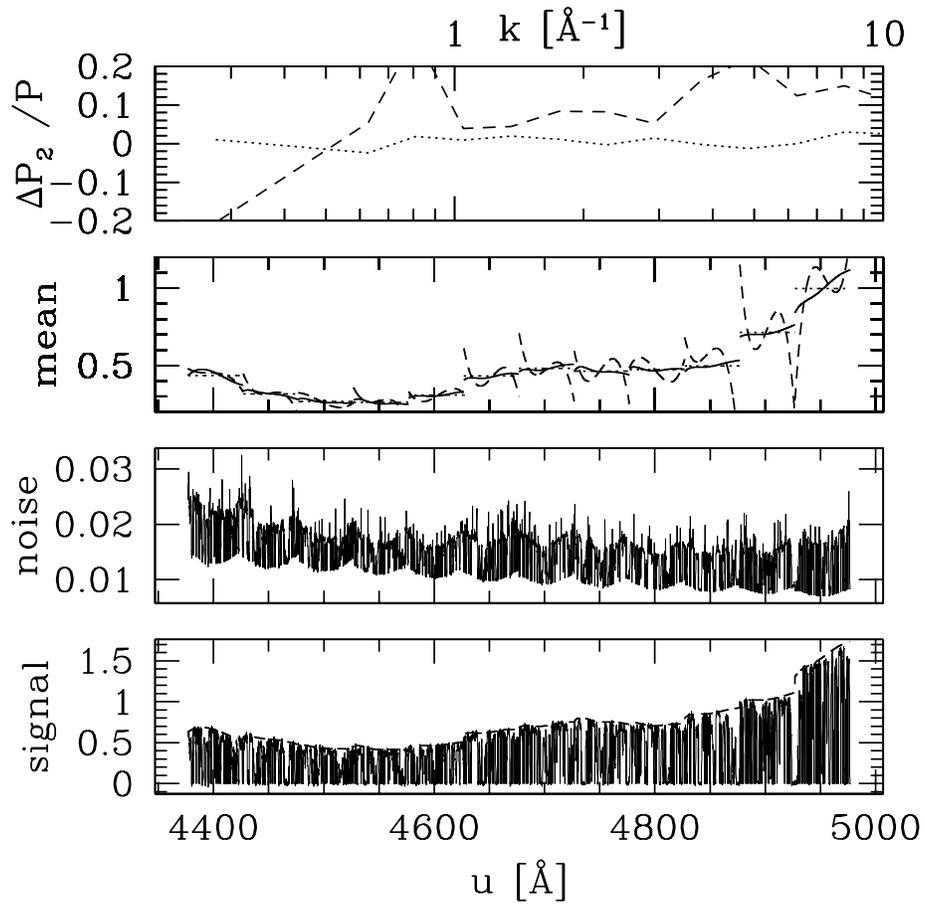,height=5.0in}}
\caption{Similar to Fig. \ref{highKeckNM} except that a relative
calibration between different echelle orders have not been done, and
the input continuum is taken from continuum-fits to an observed quasar
spectrum. The dotted line in the top two panels corresponds
to the case where the continuum is modeled as flat for each order.
The dashed line is where polynomials up to the third order are used
to fit for the mean transmission in each order. The solid line in the
second panel from the top shows the true mean transmission.} 
\label{highKeckNM.C}
\end{figure}

In Fig. \ref{highKeckNM.C}, we took the continuum-fits to an observed
quasar spectrum and use them as the input
continuum for our simulation. The simulated spectrum here represents a
case in which no relative calibration between echelle orders have been
attempted, which is often the case. This is why the
continuum in the bottom panel is broken up into 12 pieces.
The second panel from the top again illustrates the recovery of the
mean transmission: dotted line for a flat model-continuum for each
order, and dashed line for a basis of polynomials up to the third
order, also separately for each order. The solid line is the true mean
transmission. The top panel shows the accuracy of the corresponding
power spectrum estimations. The assumption of a simple flat continuum
for each order gives a power spectrum that is accurate to $\sim 1 \%$.
On the other hand, overfitting with up to third order polynomials
not only causes an under-estimation of power on large scales, but also
creates spurious power on small scales.
 
%{\bf This concerns Fig. \ref{pdsScottCont}: I will move it to a later sec%tion: 
%4.2.2; perhaps, i should actually show a figure of this.
%Or i should show a figure showing the power spectrum of
%continuum-fits shortward and longward of Ly$\alpha$. clearly
%i should also refer to a later section on uros's suggestion
%if i do decide it's a good idea. --- see pdsScottCont.plot
%and /d/lyra/lhui/LA-compute/Analyze/Test3-31-99/guide, look for
%Runs/Cont.} 

Combining Fig. \ref{highKeckNM} \& \ref{highKeckNM.C} (note that they
show the power spectrum estimation on different scales), the lessons are:
1. it is better to err on the side of underfitting the mean, which tends
to over-estimates the power on large scales, but leave the power on small
scales relatively unaffected (this relies crucially on the fact
that the continuum has fluctuations only on large scales); 2. without
sufficient prior knowledge of 
the true shape of the continuum, one can at least make conservative
statements about the small-scale power, but the large-scale power
is likely prone to systematics, unless some correction is made.

%{\bf move this to a later section: 3. one
%can use the part of the 
%observed spectrum redward of Ly$\alpha$ to gain some idea of
%what a good mean-basis is, and on what scales the continuum has
%significant fluctuations.}

One additional comment: the input continuum in Fig.
\ref{highKeckNM.C}, which is taken from fits to actual data,
certainly seems to
suggest that the observed continuum has fluctuations on scales of
an echelle order ($\sim 50 \AA$). (We will quantify this better
in \S \ref{beyond}.)
It is unclear whether this
is truly due to the intrinsic continuum, or whether it is an artifact
of imperfect blaze removal or flat-fielding (see \S \ref{describe}).
If it is the former, then $\sim 50 \AA$ represents a
fundamental limit beyond which one cannot reliably measure the
transmission power spectrum, at least not without some additional
prior knowledge of the true continuum 
(which is what we will discuss in \S \ref{beyond}). 
If it is the latter, then in 
principle one should be able to do better and extend the range of
reachable scales to larger ones.
Which is the case remains to be seen.

%combining the old highKeckNM.C_S.plot and highKeckNM.C_S2.plot
%highKeckNM.C_S.plot
%\begin{figure}[htb]
%\centerline{\psfig{figure=highKeckNM.C_S.ps,height=7.0in}}
%\caption{The bottom 2 panels are the same as in Fig.
%\ref{highKeckNM.C}.
%The solid line in the top second panel shows the true continuum level
%(as opposed to the true transmission mean level as before). 
%The dashed line is a continuum-fit to the simulated spectrum.
%In the top panel, $\Delta P/P$ denotes $(\hat P_{\rm un} - P_{\rm
%un})/P_{\rm un}$ (the un-normalized power spectrum eq. [\ref{hatP0}] \&
%[\ref{Pxidef0}]) for the upper long-dashed line; while it 
%denotes $(\hat P_{1} - P)/P$ (the normalized power spectrum, but
%estimated using the continuum-fitted data; eq. [\ref{hatPOLD}] \&
%[\ref{Pxidef}]) for the lower short-dashed line.}
%\label{highKeckNM.C_S}
%\end{figure}

{\noindent \it Test 3}

\begin{figure}[htb]
\centerline{\psfig{figure=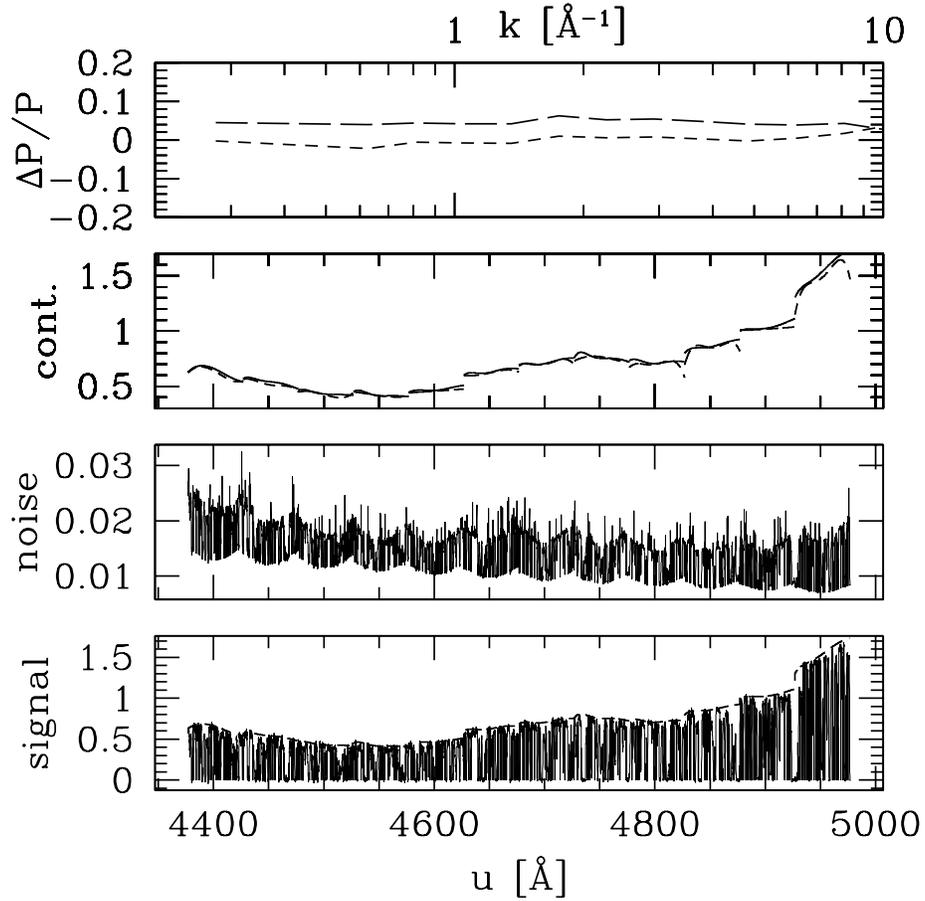,height=5.0in}}
\caption{The bottom 2 panels are the same as in Fig.
\ref{highKeckNM.C}.
The solid line in the top second panel shows the true continuum level
(as opposed to the true transmission mean level as before). 
The dashed line is a continuum-fit to the simulated spectrum.
In the top panel, $\Delta P/P$ denotes $(\hat P_{\rm un} - P_{\rm
un})/P_{\rm un}$ (the un-normalized power spectrum eq. [\ref{hatP0}] \&
[\ref{Pxidef0}]) for the upper long-dashed line; while it 
denotes $(\hat P_{1} - P)/P$ (the normalized power spectrum, but
estimated using the continuum-fitted data; eq. [\ref{hatPOLD}] \&
[\ref{Pxidef}]) for the lower short-dashed line.}
\label{highKeckNM.C_S}
\end{figure}

In Fig. \ref{highKeckNM.C_S}, we show the effect of traditional
continuum-fitting, which requires some degree of eye-balling.
The same simulated spectrum as in Fig. \ref{highKeckNM.C} is given to
an observer (one of the authors)
with no knowledge of the input continuum.
Note that the top second panel now shows the actual continuum level
rather than the mean transmission level. 
The estimated continuum actually matches the true one surprisingly
well. However, one can still see that the continuum is generally
underestimated. 
In the top panel, we show the accuracy of two different power
estimates. The long-dashed line corresponds to an estimate of the
un-normalized power spectrum as defined in eq. (\ref{Pxidef0})
(the estimator is eq.
[\ref{hatP0}]). There is clearly a $\sim 5 \%$
positive bias here because of the underestimation of the continuum. 
One way to correct for it is of course to use simulations: applying
exactly the same procedure to the observed data and the simulated
data, and see how much bias results; but
the size of the bias is likely to be model dependent.
A simple alternative way to cure this problem is to measure the normalized
power spectrum 
instead, using the continuum-fitted data, i.e. using the estimator
in eq. (\ref{hatPOLD}). This is shown with a short-dashed line.
It has an accuracy of $\sim 1 \%$, comparable to the dotted line
in Fig. \ref{highKeckNM.C}.  In view of this, it seems
by-passing continuum-fitting altogether and proceeding simply
with trend-removal is desirable.

%combined the old highKeckNM.C_S.z4.plot and highKeckNM.C_S2.z4.plot
%highKeckNM.C_S.z4.plot
%\begin{figure}[htb]
%\centerline{\psfig{figure=highKeckNM.C_S.z4.ps,height=7.0in}}
%\caption{Similar to Fig. \ref{highKeckNM.C_S} except that the
%mean transmission is lower (0.39 instead of 0.64). The solid
%and dashed line in the second panel from the top represent
%respectively the true continuum level and the continuum-fits.
%In the top panel, the upper long-dashed line shows 
%$(\hat P_{\rm un} - P_{\rm 
%un})/P_{\rm un}$ (error for the un-normalized power spectrum estimated
%from eq.
%[\ref{hatP0}]); the lower short-dashed line shows $(\hat P_{1} - P)/P$
%(error for the normalized power spectrum estimated using
%continuum-fitted data; eq. [\ref{hatPOLD}]); the dotted line shows
%$(\hat P_{2} - P)/P$ (error for the normalized power estimated
%using trend-removal with a flat trend for each echelle order; eq.
%[\ref{hatP}]).} 
%\label{highKeckNM.C_S.z4}
%\end{figure}

{\it Test 4}

\begin{figure}[htb]
\centerline{\psfig{figure=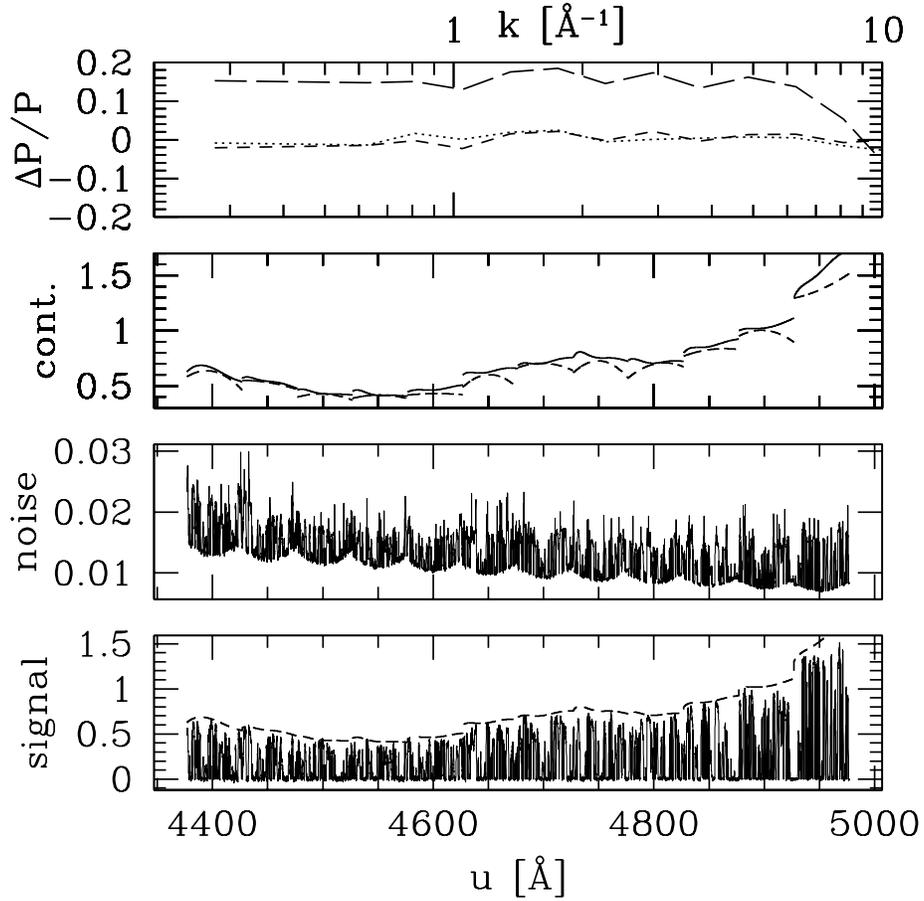,height=5.0in}}
\caption{Similar to Fig. \ref{highKeckNM.C_S} except that the
mean transmission is lower (0.39 instead of 0.64). The solid
and dashed line in the second panel from the top represent
respectively the true continuum level and the continuum-fits.
In the top panel, the upper long-dashed line shows 
$(\hat P_{\rm un} - P_{\rm 
un})/P_{\rm un}$ (error for the un-normalized power spectrum estimated
from eq.
[\ref{hatP0}]); the lower short-dashed line shows $(\hat P_{1} - P)/P$
(error for the normalized power spectrum estimated using
continuum-fitted data; eq. [\ref{hatPOLD}]); the dotted line shows
$(\hat P_{2} - P)/P$ (error for the normalized power estimated
using trend-removal with a flat trend for each echelle order; eq.
[\ref{hatP}]).} 
\label{highKeckNM.C_S.z4}
\end{figure}

The failure of traditional continuum-fitting is more dramatic
in cases where there is a lot of absorption
e.g. at high redshifts. In Fig. \ref{highKeckNM.C_S.z4} is shown a
simulated spectrum 
with the ionizing background adjusted to give a mean
transmission of 0.39, which is about the observed value
at $z = 4$ (\citenp{press93}). The continuum is more
seriously under-estimated leading to an overestimate of the
un-normalized power spectrum by $\sim 20 \%$ (the upper long-dashed
line). The normalized power spectrum, estimated either using
the continuum-fitted data (eq. [\ref{hatPOLD}]) or using directly the
trend-removed data
(eq. [\ref{hatP}]),
is much more accurately measured.

%combined the old LickNM_cont3L_S.plot and LickNM_cont3L_S2.plot
%LickNM_cont3L_S.plot
%\begin{figure}[htb]
%\centerline{\psfig{figure=LickNM_cont3L_S.ps,height=7.0in}}
%\caption{The bottom 2 panels show a low resolution (FWHM = 1.17 \AA), somewhat noisy
%simulated spectrum based on the one in Fig. \ref{NOnoise}. 
%The solid line in the second panel from the top is the true continuum 
%(not mean transmission) whereas the dashed line corresponds to a
%continuum-fit to the data in the bottom panel. 
%In the top panel, $\Delta P/P$ denotes $(\hat P_{\rm un} - P_{\rm
%un})/P_{\rm un}$ (the un-normalized power spectrum eq. [\ref{hatP0}] \&
%[\ref{Pxidef0}]) for the lower long-dashed line; while it 
%denotes $(\hat P_{1} - P)/P$ (the normalized power spectrum, but
%estimated using the continuum-fitted data; eq. [\ref{hatPOLD}] \&
%[\ref{Pxidef}]) for the upper short-dashed line.}
%\label{LickNM_cont3L_S}
%\end{figure}

{\it Test 5}

\begin{figure}[htb]
\centerline{\psfig{figure=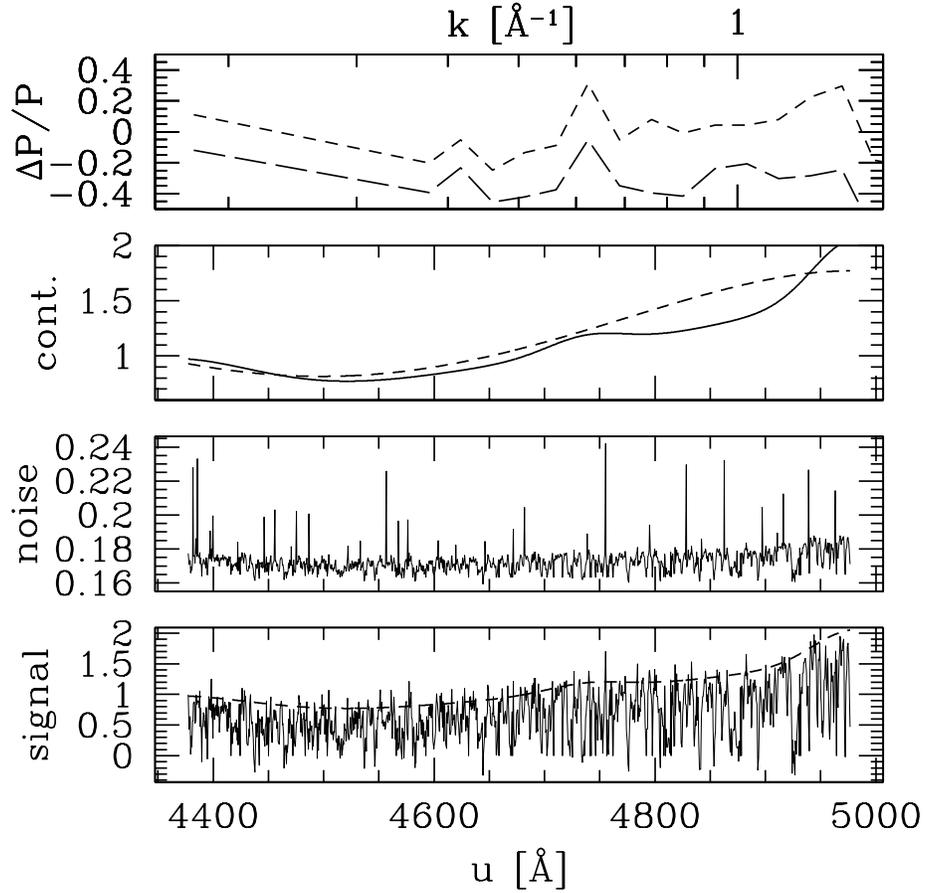,height=5.0in}}
\caption{The bottom 2 panels show a low resolution (FWHM = 1.17 \AA), somewhat noisy
simulated spectrum based on the one in Fig. \ref{NOnoise}. 
The solid line in the second panel from the top is the true continuum 
(not mean transmission) whereas the dashed line corresponds to a
continuum-fit to the data in the bottom panel. 
In the top panel, $\Delta P/P$ denotes $(\hat P_{\rm un} - P_{\rm
un})/P_{\rm un}$ (the un-normalized power spectrum eq. [\ref{hatP0}] \&
[\ref{Pxidef0}]) for the lower long-dashed line; while it 
denotes $(\hat P_{1} - P)/P$ (the normalized power spectrum, but
estimated using the continuum-fitted data; eq. [\ref{hatPOLD}] \&
[\ref{Pxidef}]) for the upper short-dashed line.}
\label{LickNM_cont3L_S}
\end{figure}

\begin{figure}[htb]
\centerline{\psfig{figure=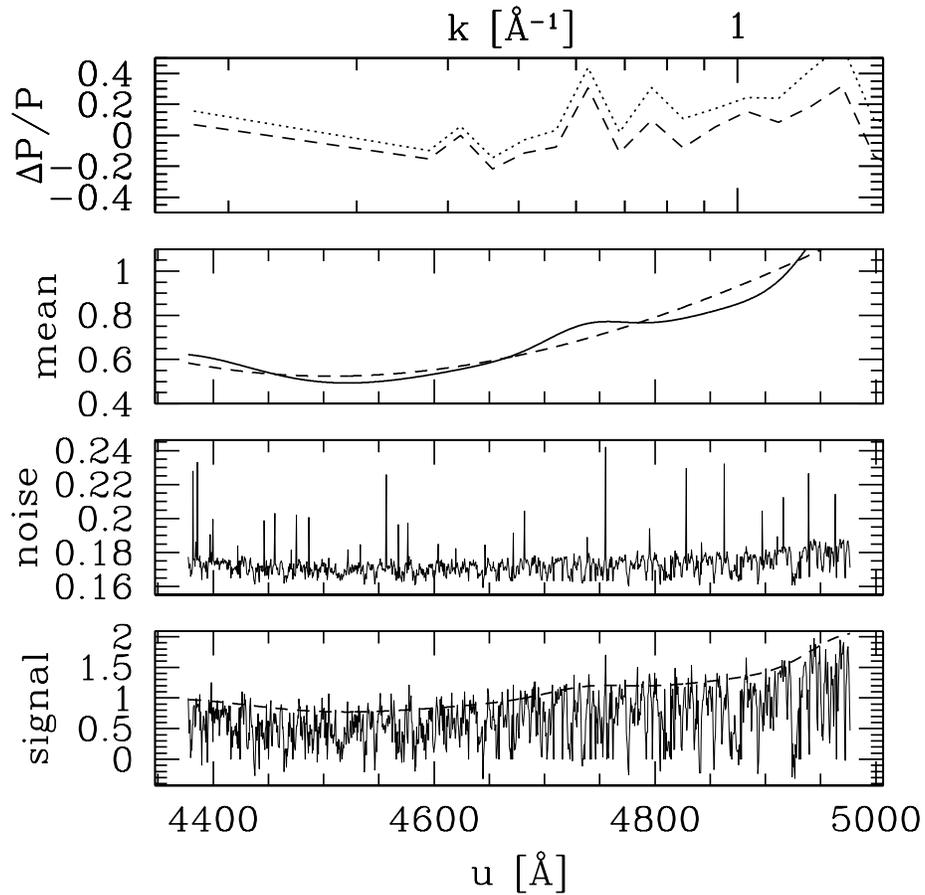,height=5.0in}}
\caption{The bottom 2 panels are the same as in Fig.
\ref{LickNM_cont3L_S}. The second panel from the top shows the
recovery
of the mean transmission assuming a model composing of
polynomials up to the third order (dashed line). The solid line
shows the true mean. The dashed line in the top panel
represents $(\hat P_{2} - P)/P$ -- the error for the normalized power
spectrum estimated using eq. (\ref{hatP}). The dotted line
shows the power spectrum estimate if shot-noise were not subtracted.}
\label{LickNM_cont3L}
\end{figure}

Another example in which traditional continuum-fitting fails
is shown in Fig. \ref{LickNM_cont3L_S}. This is based on the same
spectrum as in Fig. \ref{NOnoise}, but convolved with a Gaussian
of $1.17 \AA \,$ FWHM and with much poorer S/N compared to the simulated
spectra above. This is likely not the product of an echelle
spectrograph, hence there is no division into 12 orders.
We repeat the exercise of continuum-fitting and then power spectrum
measurement as before.
Interestingly, the significant discrete fluctuations due to the low S/N here cause an
overestimation (unlike in Test 3 and 4) of the continuum level, and so
an underestimation of
the un-normalized power spectrum. Once again, the normalized power
spectrum does not suffer from the same problem.
Note the somewhat large fluctuations of the estimated power -- this is
largely due to the high level of shot-noise.

Fig. \ref{LickNM_cont3L} shows the measurement of power spectrum
through trend removal instead. A third order polynomial is used to
estimate the mean transmission. The resulting (normalized) power 
spectrum estimate
(eq. [\ref{hatP}]) is of comparable accuracy to that using the
continuum-fitted data. We also show in the top panel as a dotted line
the power spectrum estimate without shot-noise subtraction (eq.
[\ref{wkbkuniform}]). Clearly, shot-noise introduces a bias of
the order of $10 \%$ here.
We will have some more to say about this in \S \ref{conclude}.

Tests 4 and 5 above drive home the point that the bias of an estimate
of the un-normalized power spectrum from continuum-fitted data is highly
variable. It depends on the redshift, resolution as well as S/N of the
data. 

There have been in the literature discussions of an alternative method
to normalize the quasar count: normalizing by the maximum value
of the continuum-fitted count, instead of by the mean count
(e.g. \citenp{mcdonald99}). Note that this procedure is also sensitive
to the S/N and resolution of the data. For instance, ${\rm max} (\hat N_{\rm
Q}^\alpha / \hat N_{\rm C}^\alpha ) = 1.4$ in Fig.
\ref{LickNM_cont3L_S}, while ${\rm max} (\hat N_{\rm
Q}^\alpha / \hat N_{\rm C}^\alpha ) = 1.12$ in Fig.
\ref{highKeckNM.C_S}, where
$\hat N_{\rm C}^\alpha$ is the estimated continuum -- they share 
exactly the same underlying cosmic
signal but the former has a higher level of discrete fluctuations and
poorer resolution -- 
for reference, the true maximum transmission should be 0.99.
This means one should take care to simulate the noise properties
correctly (e.g. \citenp{rauch97}).

Lastly, we should emphasize that while trend-removal seems
to be more desirable than traditional continuum-fitting for the 
particular
application here, continuum-fitting is still very useful
for other purposes, which we will discuss in \S \ref{beyond} and
\S \ref{conclude}. But a fully automated procedure for continuum-fitting
is clearly desirable.

\subsubsection{A Bonus of Trend-removal  -- Power Correction on Large Scales}
\label{beyond}

%the following represents my old attempts to get at the correction
%of large scale power; but it fails primarily because
%I don't have enough realizations of continuum and data
%
%one could still argue continuum-fitting might win on larger
%scales; might want to do some test, and argue that only as
%large as about k of 0.3 anyway...
%see /d/lyra/lhui/LA-compute/Analyze/Test3-31-99/guide, look for
%Runs/PowerLS/coeff.plot and subtract.plot

As is clear from some of the previous tests in \S \ref{continuum},
the power spectrum measured on large scales (i.e. scales comparable to
the typical scales where the continuum has fluctuations) could contain
spurious contributions from the continuum, the size of which depends
somewhat on the continuum/mean-shape-model one assumes.
The strategy adopted in the \S \ref{continuum} is a conservative one: 
assume a
model for the continuum that is as simple (or smooth) as possible, perform
trend-removal, and the resulting power spectrum would reflect the true
transmission power spectrum at least on small scales, but not necessarily on
large scales. 

\begin{figure}[htb]
\centerline{\psfig{figure=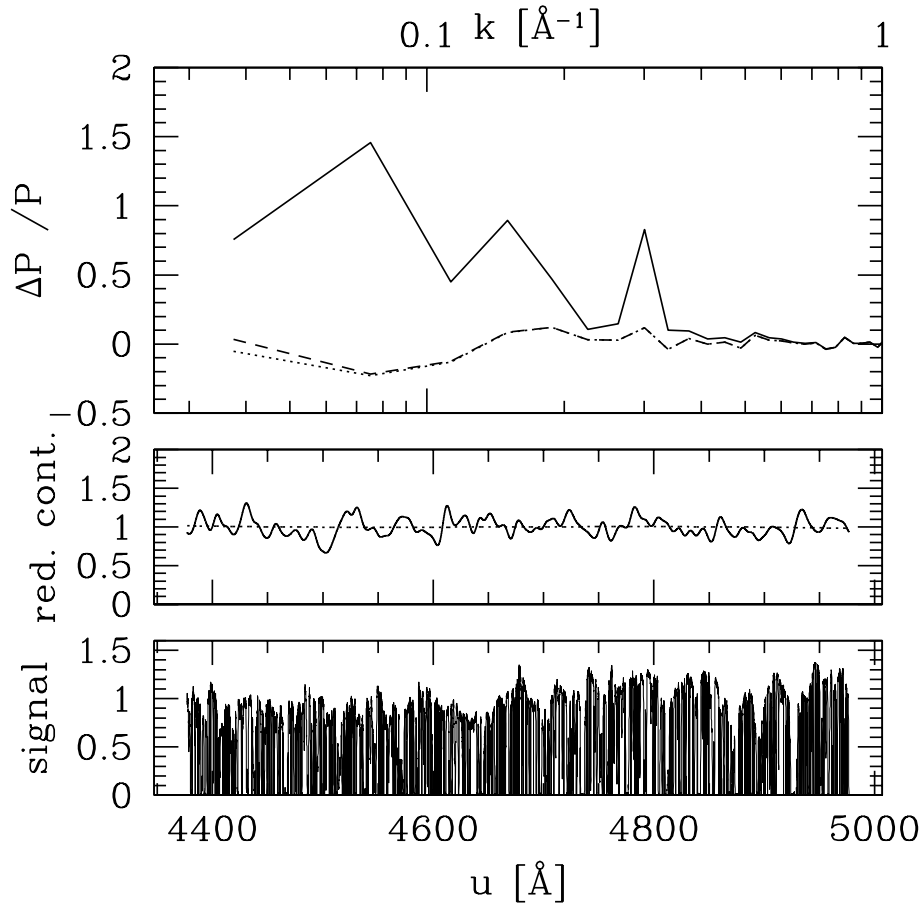,height=5.0in}}
\caption{A demonstration of power spectrum corrections on large
scales. The theoretical spectrum from Fig. \ref{NOnoise}
is multiplied by a set of 10 different continua (one of which
is shown in the middle panel), convolved with a Gaussian
of 0.125 \AA FWHM and with a small amount of noise added (similar to
Fig. \ref{highKeckNM}).
The resulting 10 simulated spectra (one of which is shown
in the bottom panel) is analyzed and the resulting power
spectrum fractional error is shown in the top panel.
Solid line shows error in the power spectrum estimate with no
corrections applied (eq. [\ref{hatP}] or [\ref{hatPd}]);
dotted line shows the error using the estimator $\hat P_4$ 
in eq. (\ref{P4}),
and dashed line shows the error using $\hat P_3$ from eq. (\ref{P3}).}
\label{combKeck}
\end{figure}

\begin{figure}[htb]
\centerline{\psfig{figure=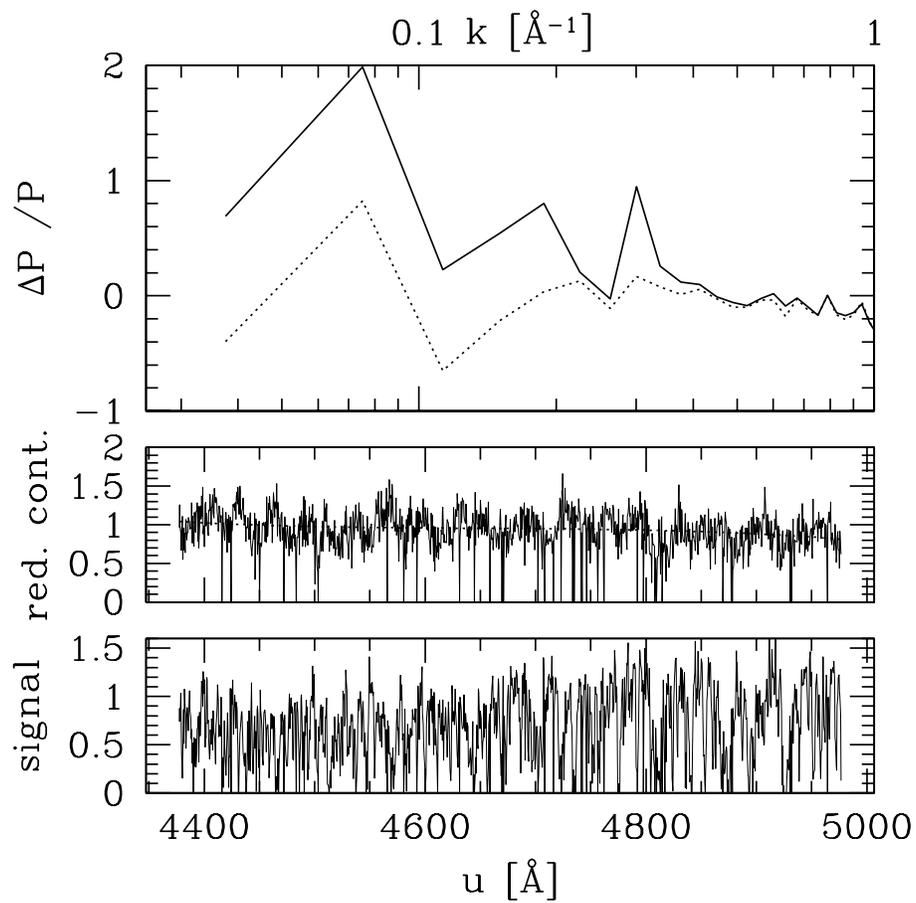,height=5.0in}}
\caption{Similar to Fig. \ref{combKeck}, except that the noise level
and resolution resemble instead those of Fig. \ref{LickNM_cont3L_S}.
Note how the noise added makes the continuum (middle panel), from which we
estimate the continuum power spectrum, quite noisy as well.
In the top panel, the solid line shows the fractional error in the
power spectrum 
estimate with no corrections applied (eq. [\ref{hatP}] or
[\ref{hatPd}]), and the dotted line shows the error using the estimator
$\hat P_4$ from
eq. (\ref{P4}).}
\label{combLick}
\end{figure}

Can we do better?
The answer is yes, under certain assumptions which we will
make explicit shortly, and it illustrates an added benefit
of trend-removal as introduced in the \S \ref{continuum}.
Readers not interested in the details can skip directly
to the end of this section where two examples of how 
our technique of power-correction
works are given (Fig. \ref{combKeck} and \ref{combLick}).

Let us start by recalling the power spectrum estimator 
in eq. (\ref{hatP}), but focusing now on the fact that
the true $\bar N^\alpha$ is unknown, and has to be estimated
using eq. (\ref{Nmeanest}), which assumes implicitly that
the true mean count obeys eq. (\ref{polyn}), which is of course
only a reasonable guess. Let us use $\hat{{\bar
N_{\rm Q}}^\alpha}$ to denote the estimated mean count, which generally
differs from the true mean count ${\bar N_{\rm Q}^\alpha}$.
We have used ${\bar N^\alpha}$ somewhat sloppily before
when we really meant $\hat{{\bar N_{\rm Q}}^\alpha}$ e.g. eq. 
[\ref{Nmeanest}].
In other words, eq. [\ref{Nmeanest}] should be more accurately written as
\begin{equation}
\hat {\bar N_{\rm Q}^\alpha} = \sum_\beta L^{\alpha\beta} \hat N_{\rm Q}^\beta \, \, , \, \, 
{\bf L} \equiv {\bf p^T (p p^T)^{-1} p}
\label{Nmeanest2}
\end{equation}
where ${\bf p}$ represents the basis functions.
Similarly, the estimator for the power spectrum in eq. (\ref{hatP}) 
should be more accurately written as:
\begin{mathletters}
\label{hatPd}
\begin{eqnarray}
\label{hatPPd}
&& \hat \xi_{2} (u) = \sum_{\alpha,\beta} w^{\alpha\beta} (u)
\hat d_f^\alpha \hat d_f^\beta \, \, , \, \,
\hat P_{2} (k) = \sum_{\alpha,\beta} w^{\alpha\beta} (k)
\hat d_f^\alpha \hat d_f^\beta - b (k) \\
\label{hatdeltafd}
&& \hat d_f^\alpha \equiv (\hat N_{\rm Q}^\alpha 
-\hat {\bar N_{\rm Q}^\alpha}) / \hat {{{\bar N}_{\rm Q}}^\alpha} = 
\sum_\gamma D^{\alpha\gamma} \hat N_{\rm Q}^\gamma / 
\sum_\eta L^{\alpha\eta} \hat N_{\rm Q}^\eta \, \, , \, \, 
D^{\alpha\gamma} \equiv \delta^{\alpha\gamma} - L^{\alpha\gamma}
\end{eqnarray}
\end{mathletters}

We now assume the following quantities are small:
$\hat d_f^\alpha$ and $(\sum L^{\alpha\gamma} \hat N_{\rm Q}^\gamma 
- \bar N_{\rm Q}^\alpha)/\bar N_{\rm Q}^\alpha$. 
The second quantity tells us how
far off our estimate of the mean is from the true mean,
while the first contains contributions both from the fluctuation
in transmission and from the second quantity.
Therefore, putting eq. (\ref{hatdeltafd}) and eq. (\ref{hatPPd}) 
together, the lowest order contributions to the expectation
value of the estimator $\hat P_2 (k)$ are:
\begin{mathletters}
\label{expectP20}
\begin{eqnarray}
\label{expectP2}
\langle \hat P_2 (k) \rangle
&=& \sum_{\alpha\beta\gamma\eta} w^{\alpha\beta} (k) \langle D^{\alpha\gamma} D^{\beta\eta} {{\bar N_{\rm Q}^\gamma}{\bar N_{\rm Q}^\eta} \over {{\bar N_{\rm Q}^\alpha}{\bar N_{\rm Q}^\beta}}} \rangle (1 + \langle \hat \delta_f^\gamma \hat \delta_f^\eta \rangle) \\ \nonumber
&=& P_{\rm C} (k) + \int {dk'\over {2\pi}} P(k') G_n (k,k') \\
\label{expectP2b}
P_{\rm C} (k) &\equiv& \sum_{\alpha\beta\gamma\eta} w^{\alpha\beta} (k) \langle
D^{\alpha\gamma} D^{\beta\eta} {{\bar N_{\rm Q}^\gamma}{\bar N_{\rm Q}^\eta} \over {{\bar N_{\rm Q}^\alpha}{\bar N_{\rm Q}^\beta}}} \rangle \\ 
\label{expectP2c}
G_n (k,k') &\equiv& \sum_{\alpha\beta\gamma\eta} w^{\alpha\beta} (k) e^{ik'(u^\gamma - u^\eta)}
\langle D^{\alpha\gamma} D^{\beta\eta} {{\bar N_{\rm Q}^\gamma}{\bar N_{\rm Q}^\eta} \over {{\bar N_{\rm Q}^\alpha}{\bar N_{\rm Q}^\beta}}} \rangle
\, , \\
\end{eqnarray}
\end{mathletters}
where we have retained the old definition of $\hat \delta_f^\gamma$
as $(\hat N_{\rm Q}^\gamma - \bar N_{\rm Q}^\gamma)/\bar 
N_{\rm Q}^\gamma$ (eq. [\ref{hatP}]). The above gives
an idea of how biased the estimator $\hat P_2 (k)$ is.
Note that we have used $\langle \, \rangle$ here to include,
in addition to the ensemble averaging as explained in \S \ref{notation}, 
an averaging over the ensemble of possible continua (which changes
$\bar N_{\rm Q}^\alpha$ because it is directly 
proportional to the continuum count $N_{\rm C}^\alpha$).
We have assumed the fluctuations in the continuum are uncorrelated
with fluctuations in the cosmic signal $\hat \delta_f^\gamma$.
We have also ignored the shot-noise contributions
(e.g. $b (k)$) and will continue to do so for the
rest of this section, because the scales where the continuum contamination
could be a problem are typically large enough that shot-noise
is subdominant.

The term $P_{\rm C} (k)$ can be viewed as the power spectrum
of the continuum fluctuation. This is fluctuation in the
sense of $D^{\alpha\gamma} \bar N_{\rm Q}^\gamma = 
\bar N_{\rm Q}^\alpha - L^{\alpha\gamma} \bar N_{\rm Q}^\gamma$.
This fluctuation would vanish if our trend-removal
procedure were so accurate that the continuum shape is exactly 
reproduced. 
%We expect that $\epsilon (k)$ could be comparable to 
%$P (k)$ on 
%sufficiently large scales.
The term $G_n (k,k')$ is the effective window
function, replacing the one in eq. (\ref{Gwindow}),
which does not take into account the error involved in 
trend-removal. The desirable normalization condition 
$\int G_n (k,k') dk'/(2\pi) = 1$ no longer holds with
the choice of $w^{\alpha\beta}$ in eq. (\ref{wkuniform}).
We have instead
\begin{equation}
\int G_n (k,k') {dk'\over 2\pi} = {1\over {\cal N}} 
\sum_{\alpha\beta\gamma} R^{\alpha\beta} (k) 
D^{\alpha\gamma} D^{\beta\gamma} 
{(\bar N_{\rm Q}^\gamma)^2 \over
{{\bar N_{\rm Q}^\alpha} {\bar N_{\rm Q}^\beta}}} \equiv 
1 + \epsilon_G (k)
\label{epsilonG}
\end{equation}
where $R^{\alpha\beta} (k)$ is defined in eq. (\ref{wkuniform}).

Assuming for now $P_{\rm C} (k)$ and $\epsilon_G (k)$ can be measured
from a suitable ensemble of continua, we propose the following
alternative estimator to $\hat P_2 (k)$, which removes the
bias due to continuum contamination:
\begin{equation}
\hat P_3 (k) = [\hat P_2 (k) - P_{\rm C} (k)]/[1 + \epsilon_G (k)]
\label{P3}
\end{equation}

The above gives us an unbiased estimate of the {\it windowed} power
spectrum. The window is effectively $G_n (k,k')/[1+\epsilon_G (k)]$
which has the desirable normalization.
We will not attempt further improvements such as
deconvolution in this paper.

A useful alternative estimator, in cases where $P_{\rm C}$ dominates
the bias in $\hat P_2 (k)$, is
\begin{equation}
\hat P_4 (k) = \hat P_2 (k) - P_{\rm C} (k)
\label{P4}
\end{equation}
The above estimator gets rid of most of the bias in the
estimator $\hat P_2 (k)$ if $P_{\rm C} (k)/P (k) \gg \epsilon_G (k)$.
An interesting corollary is that, under such a condition,
the bias in $\hat P_2 (k)$ is positive since $P_{\rm C} (k)$ is
positive definite. Needless to say, this statement breaks down
if $P_{\rm C} (k)$ is not the dominant source of bias, or
if the fractional error in the mean-estimation is large
(see e.g. Fig. \ref{highKeckNM}).

It is interesting to compare our derivation above with
the well-known one for the integral constraint bias
in galaxy surveys
(s.g. \citenp{peebles80,ls93,bernstein94,tegmark98}).
The integral constraint arises from the fact that the
mean density of a galaxy survey has to be estimated
from the same survey from which one is also measuring
the power spectrum. The fact that the power spectrum
estimator involves a non-trivial nonlinear combination
of the data gives rise to a bias (see \citenp{hg99}),
quite analogous to our derivation here.
However, in the standard derivations,
it is assumed the shape of the mean density is known
(often taken to be uniform), and therefore 
$P_{\rm C} (k)$ effectively vanishes, whereas $\epsilon_G (k)$
can be non-negligible on scales comparable to the size of the survey,
but is otherwise small.
The reader is referred to Bernstein \cite*{bernstein94} and
Hui \& Gazta\~naga \cite*{hg99} for discussions on higher order
contributions to the integral constraint.

How should one estimate $P_{\rm C} (k)$ and $\epsilon_G (k)$?
Given an ensemble of continua (with counts represented by
$N_{\rm C}^\alpha$), our procedure is to replace
$\bar N_{\rm Q}^\alpha$, which appears in the definitions
of $P_{\rm C} (k)$ and $\epsilon_G (k)$ (eq. [\ref{expectP2b}],
[\ref{epsilonG}]) with $N_{\rm C}^\alpha$, and compute
the corresponding ensemble averages.
Note that $\bar N_{\rm Q}^\alpha = N_{\rm C}^\alpha 
\langle e^{-\tau} \rangle$, but $\langle e^{-\tau} \rangle$, which
is taken to be constant over the finite redshift range of interest,
is divided out
in the relevant definitions of $P_{\rm C} (k)$ and $\epsilon_G (k)$.

The hard question is of course how to obtain a suitable ensemble
of continua.
The first thing one might try is to measure the power spectrum of
the continuum-fits (i.e. $P_{\rm C} (k)$, or more generally, both
$ P_{\rm C} (k)$ and
$\epsilon_G (k)$) from exactly
the same regions from which one attempts to measure the transmission
power spectrum. While this can give us a crude
idea of how significant the continuum power spectrum is,
it is not entirely satisfactory because part
of what has been ascribed to the continuum might actually
be large scale fluctuations in the cosmic signal $\delta_f^\alpha$
that we are after, or vice versa.

The second option that comes to mind is to measure the
continuum power spectrum from regions where the continuum determination
is relatively secure. Two possibilities are
{\bf A.} low redshift quasar spectra
where the forest is much less dense, and {\bf B.} 
regions of spectra which
lie redward of Ly$\alpha$ emission. The working hypothesis is
that the continuum power spectrum in these two regions
is the same as, or at least similar to, the one in the region where 
we attempt to
estimate the transmission power spectrum (the forest of interest).
There is no guarantee that the hypothesis is valid. 
For instance, regarding possibility {\bf A}, 
the continuum power could systematically
evolve with redshift. In fact it probably does: assuming
that the statistical properties of the quasar continuum {\it in rest frame} do not evolve with redshift, the observed
continuum power would evolve as $P_{\rm C} [k, z_1] =
P_{\rm C} [k(1+z_0)/(1+z_1), z_0]$. One could in principle
constrain such redshift evolution with a sufficiently large
sample of low redshift quasar spectra. 
Regarding possibility {\bf B}, it is not unreasonable
to expect that the continuum power is higher on
the red side compared to the blue side, because
there are generally more broad emission lines on the
red side (see e.g. \citenp{peterson97,blandford90}; 
see below for caveats and
a counter example, however). An upper bound on the blue
continuum power is by itself interesting because one
can then obtain a conservative estimate of how much
spurious power is introduced by the continuum into one's
forest power measurements. Furthermore, systematic differences
between the red and blue continuum power can be
studied and quantified with a sufficiently large sample
of low redshift quasars. 

%see pdsScottCont3.plot (the old version used pdsScottCont.plot)
%and /d/lyra/lhui/LA-compute/Analyze/Test3-31-99/guide, look for
%Runs/Cont
%pdsScottCont2.plot (the old version used pdsScottCont.plot)
%QSO 1157+3143  
%see Workspace/lhui/Fermi/honus/lhui/Theory/Da/Corr/9_4/Another
%\begin{figure}[htb]
%\centerline{\psfig{figure=pdsScottCont.ps,height=7.0in}}
%\caption{The bottom two panels show continuum-fits taken
%from the red (bottom) and blue (second from bottom) sides
%of Ly$\alpha$ emission from the spectrum of a 
%quasar at redshift $z = 3$. The top panel shows
%the power spectra of the red and blue continua (eq. [\ref{expectP2b}]) 
%with
%solid and dotted lines respectively. They are very close
%to each other, the panel below it shows their fractional
%difference $[P_{\rm C} ({\rm blue}) - P_{\rm C} ({\rm red})] /
%P_{\rm C} ({\rm red})$. Also shown in the top panel is the
%quantity $\epsilon_G (k)$ (eq. [\ref{epsilonG}]), 
%with solid and dotted lines
%denoting its values on the red and blue sides. They
%differ by only a few percent.}
%\label{pdsScottCont}
%\end{figure}

\begin{figure}[htb]
\centerline{\psfig{figure=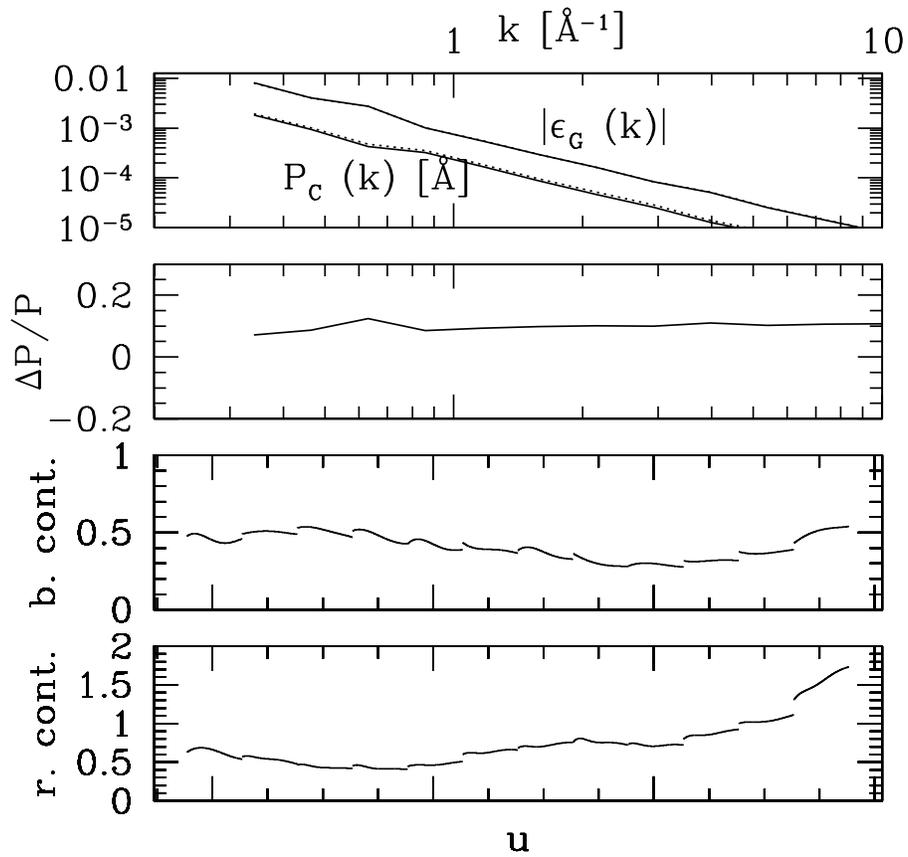,height=5.0in}}
\caption{The bottom two panels show continuum-fits taken
from the red (bottom) and blue (second from bottom) sides
of Ly$\alpha$ emission from the spectrum of a 
quasar at redshift $z = 3$. The top panel shows
the power spectra of the red and blue continua (eq. [\ref{expectP2b}]) 
with
solid and dotted lines respectively. They are very close
to each other, the panel below it shows their fractional
difference $[P_{\rm C} ({\rm blue}) - P_{\rm C} ({\rm red})] /
P_{\rm C} ({\rm red})$. Also shown in the top panel is the
quantity $\epsilon_G (k)$ (eq. [\ref{epsilonG}]), 
with solid and dotted lines
denoting its values on the red and blue sides. They
differ by only a few percent.}
\label{pdsScottCont}
\end{figure}

In Fig. \ref{pdsScottCont}, we show the continuum power spectrum
measured from the continuum estimates on both sides of the
Ly$\alpha$ emission of a quasar at z=3 (QSO 1157+3143).
%QSO 1157+3143  
%see /workspace/lhui/Fermi/honus/lhui/Theory/Da/Corr/9_4/Another
% or /d/honus/lhui/Theory/Da/Corr/9_4/Another
The continuum estimates
are shown in the bottom two panels. After fitting a flat mean
to each echelle order, we compute the continuum power spectrum
just as if this were the forest, and the results 
from the red side and blue side are shown as solid and
dotted lines respectively in the top panel. The two power spectra look
similar. However, we emphasize that because of the lack of small scale
power in the continuum, most of the power on small scales
($k \, \approxgt \, 1 \AA^{-1}$) that we see in 
Fig. \ref{pdsScottCont} is likely aliased from large scales.
We will not attempt to perform a deconvolution to obtain
the true small scale power; it suffices to note here that the
true small scale power can only be smaller than what is shown in
the figure.
Also shown in the top panel is $\epsilon_G (k)$, on
both sides of Ly$\alpha$, which are basically indistinguishable
from each other. Note that $\epsilon_G (k) \ll 1$.
The second panel from the top shows the fractional difference
between $P_{\rm C}$ from the red and blue sides,
which is about $10 \%$, with the blue continuum power
systematically higher than the red one.
The results here, though drawn from admittedly a very small sample,
are quite interesting for several reasons.

\begin{itemize}

\item The excess of the blue continuum power spectrum over
the red one is consistent with the following hypothesis: that some
of the fluctuations in the forest have been wrongly assigned
to the continuum during the continuum-fitting process on
the blue side. In other words, the true blue continuum power spectrum
should be lower than the dotted line in the top panel of 
Fig. \ref{pdsScottCont}. An upper bound on the true blue continuum power
spectrum is already very useful. One can use it to quantify how
much, and on what scales, one should worry about spurious continuum
power introduced into estimation of the transmission power.
One can compare Fig. \ref{pdsScottCont} with the theoretical
expectation in Fig. \ref{NOnoise}, and see that the spurious power
must be negligible for $k \approxgt 0.3 \AA^{-1}$. 
This explains why the determination of the transmission power
spectrum from both the continuum-fitted data or the trend-removed
data is very accurate in examples like Fig. \ref{highKeckNM.C_S.z4},
as long as one considers the {\it normalized} power.
Unfortunately, the pieces of continuum we examine are not long
enough to yield useful information on larger scales or smaller $k$'s.
If one takes a crude extrapolation, the continuum power spectrum
(or more accurately its upper bound)
might become non-negligible compared to the transmission power spectrum 
at $k \sim 0.1 \AA^{-1}$. However, one must keep in mind that
the theoretical transmission power spectrum in
Fig. \ref{NOnoise} is likely underestimated at small $k$'s because
the simulation lacks large scale power. Nonetheless, there should
be a genuine flattening of the transmission power spectrum at
large scales. In any case, the first point to bear in mind 
is that an upper bound on the continuum power spectrum is
useful as a conservative estimate of the possible spurious power.
\footnote{One should be aware of a possible pitfall of the
above argument, however. It is not impossible that
the opposite can happen, that one underfits the blue continuum,
and ends up underestimating the blue continuum power.
This is probably not the case here, where the data from which
the blue continuum is estimated have high signal-to-noise and
resolution (similar to the simulated spectrum
in Fig. \ref{highKeckNM.C_S}, where it can be seen that
the continuum fit tends to have features that follow the
forest). Underfitting the blue continuum
is more likely for low resolution data, although even there,
the situation is not clear: underfitting would result
in underestimation of the continuum power on small scales,
but not necessarily on large scales. Obviously, more tests
are needed.}

\item Further, one can test the hypothesis that the excess in blue 
continuum power is due to contamination from the forest:
if this is true, one expects the red and blue continuum power spectra
to converge,
as one goes to lower redshift quasars, because presumably,
the blue continuum power spectrum should be less affected by
the forest at lower redshifts. Even if their difference does not
converge to zero (as suggested by the larger
number of broad emission lines on the red side), 
but to some small but finite value, this
is still a useful exercise because it gives us an idea of
how different the red and blue continuum power spectra can be.
If we can determine the blue continuum power spectrum to
an accuracy of $10 \%$ say, and use this to correct for the 
transmission power spectrum on large scales, this is already a 
significant improvement
over not correcting for the large scale power, or 
simply throwing away the information on large scales
altogether. For instance, if the blue continuum power
does become comparable with the transmission power at $k \sim 0.1 \AA^{-1}$, not subtracting off the spurious power would result in a fractional
error of $100 \%$, while subtracting off an approximate blue continuum
power accurate to $10 \%$ reduces the error by an order
of magnitude.

\end{itemize}

Obviously, more testing using observed data is warranted,
particularly on the estimation of red and blue
continuum power as a function of redshift. This will be carried
out in a separate paper. 
One natural question that might occur to the reader is
whether a universal continuum power spectrum actually exists,
given the large observed variations in the continuum
from one quasar to another. 
It suffices to note that given an ensemble,
the averaged power spectrum is {\it always} a well-defined
quantity. The tricky part is to make sure the ensemble
from which one estimates the continuum power spectrum
has the same averaged continuum power spectrum as the ensemble
of continua in the forest regions of interest.
As a simple example: one might want to make sure the
same proportion of radio-loud quasars are included
in both ensembles. This is probably desirable if one uses the
working hypothesis that low redshift blue continuum power
is similar to high redshift blue continuum power, as suggested
above. Alternatively, if the hypothesis that blue and red continuum
power spectra resemble each other irrespective of redshift
turns out to be a reasonable one, the simplest way to make
sure one has the right ensemble is to use both sides 
of Ly$\alpha$ for any give quasar: use the blue side for
its forest, and the red side for its continuum.

With all of the above caveats in mind, let us illustrate
the technique of power correction with two idealized examples,
where it is assumed the right ensemble of continua is 
in hand.

%combKeck.plot
%\begin{figure}[htb]
%\centerline{\psfig{figure=combKeck.ps,height=7.0in}}
%\caption{A demonstration of power spectrum corrections on large
%scales. The theoretical spectrum from Fig. \ref{NOnoise}
%is multiplied by a set of 10 different continua (one of which
%is shown in the middle panel), convolved with a Gaussian
%of 0.125 \AA FWHM and with a small amount of noise added (similar to
%Fig. \ref{highKeckNM}).
%The resulting 10 simulated spectra (one of which is shown
%in the bottom panel) is analyzed and the resulting power
%spectrum fractional error is shown in the top panel.
%Solid line shows error in the power spectrum estimate with no
%corrections applied (eq. [\ref{hatP}] or [\ref{hatPd}]);
%dotted line shows the error using the estimator $\hat P_4$ 
%in eq. (\ref{P4}),
%and dashed line shows the error using $\hat P_3$ from eq. (\ref{P3}).}
%\label{combKeck}
%\end{figure}

In Fig. \ref{combKeck}, we show in the bottom panel a simulated
spectrum with a somewhat unusual continuum (middle panel) with
a fair amount of fluctuations.
We generate a set of 10 different continua and impose each on
our underlying cosmic absorption to obtain a set of 10 different
simulated spectra (only one of which is shown in the figure). We
compute the power spectrum
using $P_2 (k)$ as in eq. (\ref{hatPd}). The resulting fractional
error from the true transmission power spectrum is shown
as a solid line in the top panel. There is clearly a lot of spurious
power on large scales due to the imperfectly estimated mean count,
which reflects the wild fluctuations in the continuum.
We then apply the power spectrum corrections: the dotted line shows
$\hat P_3 (k)$ from eq. (\ref{P3}) while the dashed line shows 
$\hat P_4 (k)$ from eq. (\ref{P4}).
One can see that subtracting the continuum power spectrum $P_{\rm C} (k)$
alone removes most of the
spurious power.

To make the example realistic, we have
multipled the continua in the forest region by a power law that
goes like $(u^\alpha)^{0.96}$ (i.e. the 'blue' continuum), and similarly 
multiplied
the continua from which we actually estimate the continuum power
by a power law of $(u^\alpha)^{-0.01}$ (i.e. the hypothetical
'red' continuum). This is meant to mimic a possible turn-over
of the quasar continuum around Ly$\alpha$ (see e.g.
\citenp{zheng98} for evidence of a turn-over around Ly$\beta$).
We have in mind
a situation in which the continuum power spectrum $P_{\rm C}$ is
estimated from the red side of Ly$\alpha$. Clearly, the fact
that the mean trends on the blue and red are different
does not present an obstacle.

%combLick.plot
%\begin{figure}[htb]
%\centerline{\psfig{figure=combLick.ps,height=7.0in}}
%\caption{Similar to Fig. \ref{combKeck}, except that the noise level
%and resolution resemble instead those of Fig. \ref{LickNM_cont3L_S}.
%Note how the noise added makes the continuum (middle panel), from which we
%estimate the continuum power spectrum, quite noisy as well.
%In the top panel, the solid line shows the fractional error in the
%power spectrum 
%estimate with no corrections applied (eq. [\ref{hatP}] or
%[\ref{hatPd}]), and the dotted line shows the error using the estimator
%$\hat P_4$ from
%eq. (\ref{P4}).}
%\label{combLick}
%\end{figure}

In Fig. \ref{combLick} we show a similar version of the above,
but with much noisier data and poorer resolution, and
a mean power-law of $(u^\alpha)^{1.5}$ and a mean
of $(u^\alpha)^{-0.9}$ have been imposed on
the continua on the blue and red sides respectively.
The same technique works here as well.

One last point we should make: when the quantities $P_{\rm C} (k)$
and $\epsilon_G (k)$ are estimated from some
ensemble of continua, they in general receive shot-noise
contributions. We have ignored shot-noise here, assuming
the scales where power correction is most interesting
are sufficiently large that shot-noise is unimportant.
This should be checked on a case by case basis.

\subsubsection{Gaps and Metal Absorption Lines}
\label{gaps}

There are at least two other possible sources of systematic errors in
addition to that due to continuum-fitting. 
Gaps are quite common in observed spectra due to defects in the CCD,
incomplete spectral coverage, or cosmic ray hits. Fortunately, since
they are at known
locations, we can either consider only those parts of the spectrum
that are
between the gaps (for instance when the gaps are large), or
interpolate to fill in the gaps (for instance when the gaps are
small). The latter is what we have implicitly done in all of the tests
mentioned in \S \ref{continuum}, where $3 \%$ of the pixels
are assumed discarded because of cosmic-ray hits. The hits are
typically one to a few pixels wide, and we simply fill them in by
linearly
interpolating the counts from neighboring pixels. Clearly, we can
recover the power 
spectrum quite well in spite of the need to interpolate. 

%this combined the old highKeckM.C.plot and highKeckMgap.C.plot
%highKeckM.C.plot
%the metal lines are taken originally from
%/work/lhui/Workspace/Fermi/honus/lhui/Theory/Da/Corr/7_31
%\begin{figure}[htb]
%\centerline{\psfig{figure=highKeckM.C.ps,height=7.0in}}
%\caption{The bottom panel shows a simulated spectrum with resolution
%and S/N similar to Fig.
%\ref{highKeckNM.C}, except that metal absorption lines
%shown in the second panel have been added. The y-axis of the second
%panel is $e^{-\tau}$ where $\tau$ is the optical depth due to metal
%absorption. The seond panel on the top shows the true mean
%transmission and the recovered mean
%transmission assuming a flat trend for each order (dotted line).
%The dotted line in the top panel is $(\hat P_{2} - P)/P$ for
%the case where no attempt is made to cut out the metal lines;
%the dashed (solid) line is the same fractional error for the
%normalized power spectrum for the case where all metal lines
%with $\tau > 1$ ($\tau > 0.4$) are discarded and the corresponding gaps
%filled in via interpolation.}
%\label{highKeckM.C}
%\end{figure}

\begin{figure}[htb]
\centerline{\psfig{figure=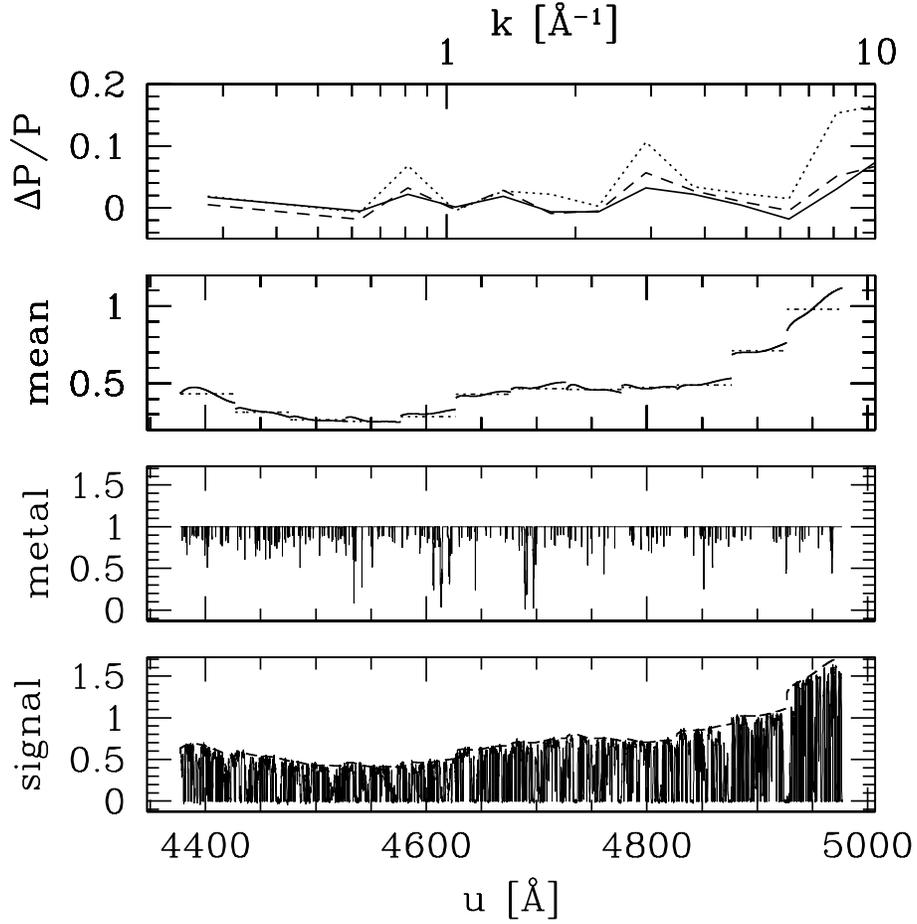,height=5.0in}}
\caption{The bottom panel shows a simulated spectrum with resolution
and S/N similar to Fig.
\ref{highKeckNM.C}, except that metal absorption lines
shown in the second panel have been added. The y-axis of the second
panel is $e^{-\tau}$ where $\tau$ is the optical depth due to metal
absorption. The seond panel on the top shows the true mean
transmission and the recovered mean
transmission assuming a flat trend for each order (dotted line).
The dotted line in the top panel is $(\hat P_{2} - P)/P$ for
the case where no attempt is made to cut out the metal lines;
the dashed (solid) line is the same fractional error for the
normalized power spectrum for the case where all metal lines
with $\tau > 1$ ($\tau > 0.4$) are discarded and the corresponding gaps
filled in via interpolation.}
\label{highKeckM.C}
\end{figure}

A more challenging problem is possible systematics due to the presence
of metal
absorption lines. Shown in the bottom panel of Fig. \ref{highKeckM.C}
is a simulated spectrum with resolution and S/N very similar to that
of Fig. \ref{highKeckNM.C} except that metal absorption lines as
shown in the panel above have been added on top of
the Ly$\alpha$ forest. This list of lines is taken from a quasar
spectrum redward of Ly$\alpha$ (HS 1103+6416, z = 2.19).
The mean transmission is estimated by
assuming a flat trend for each echelle order as before. 
What is interesting is the dotted line in the top panel, demonstrating
the creation of spurious power by the metal lines. 
The dashed / solid line shows fractional error in the power spectrum
estimate if 
all metal lines with $\tau > 0.4$ / $\tau > 1$ are assumed
``detected'', and 
therefore cut-out and treated as gaps as before (i.e. interpolated
across). Such a procedure seems to eliminate much of the spurious
power. In practice, sufficiently strong metal lines should be
identifiable by their narrow widths.

%this combined the old highKeckM.C.plot and highKeckMgap.C.plot
%LickM_cont3L.plot
%\begin{figure}[htb]
%\centerline{\psfig{figure=LickM_cont3L.ps,height=7.0in}}
%\caption{Power spectrum measurements from low resolution and noisy
%quasar spectra (similar to Fig. \ref{LickNM_cont3L_S}), with (dotted)
%and without (dashed) metal contamination.} 
%\label{LickM_cont3L}
%\end{figure}

\begin{figure}[htb]
\centerline{\psfig{figure=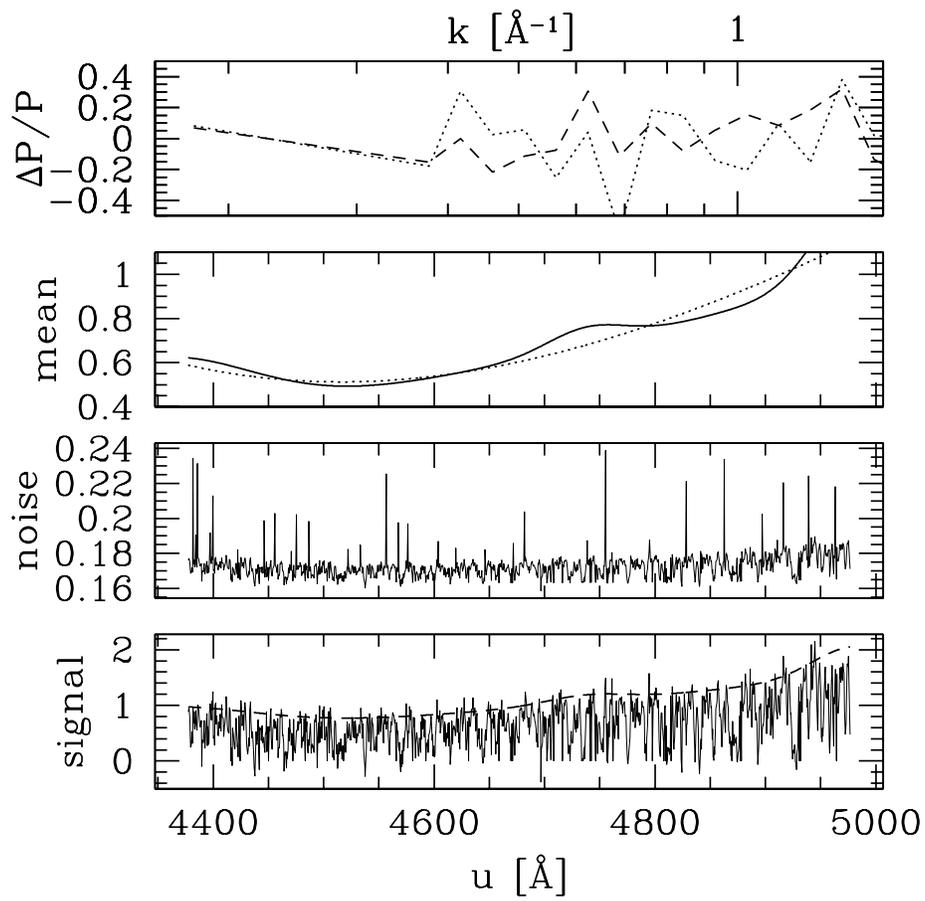,height=5.0in}}
\caption{Power spectrum measurements from low resolution and noisy
quasar spectra (similar to Fig. \ref{LickNM_cont3L_S}), with (dotted)
and without (dashed) metal contamination.} 
\label{LickM_cont3L}
\end{figure}

Fig. \ref{LickM_cont3L} shows that metal absorption lines
in data with lower resolution and poorer signal to noise
have a relatively small effect on the accuracy of the
power spectrum estimation.

\subsection{Random Errors: Shot-noise-bias, Variance and Minimum
Variance Weighting}
\label{random}

Random errors arise firstly from (cosmic) sampling fluctuations, 
and secondly from electron/photon counting, which
can be traced to fluctuations in the intrinsic quasar
counts, the sky counts and the readout (see \S \ref{describe}).
We will summarily refer to the latter as shot-noise. Shot-noise
affects two aspects of power spectrum estimation.

First, shot-noise introduces a bias which has to be subtracted off.
This is the term $b (k)$ in eq. (\ref{bkuniform}). We will
give here a more general expression for $b (k)$ suitable for different
weightings ($w^{\alpha\beta}$). As we have demonstrated in
Fig. \ref{LickNM_cont3L}, shot-noise-bias subtraction can
be important for low S/N data. We will return to this point
in \S \ref{conclude}.

Second, shot-noise, together with cosmic fluctuations, determines
the variance of the power spectrum estimate.
We will give the expression for the variance in this section, and then
address the question of how to best combine data with different
levels of S/N to minimize the variance.

The power spectrum estimator we will focus on is given in eq.
(\ref{hatP}). It is assumed trend removal as explained in \S
\ref{continuum} has been performed. We ignore uncertainties due
to the unknown continuum in this section.
Here we do not limit ourselves to the choice
of uniform weighting (eq. [\ref{wkbkuniform}]) as we have
done so far. Let us rewrite $w^{\alpha\beta} (k)$ in eq. (\ref{hatP}) as
\begin{equation}
w^{\alpha\beta} (k) = \bar w^{\alpha\beta} (k) R^{\alpha\beta} (k)
\label{barw}
\end{equation}
where $R^{\alpha\beta} (k)$ is given in eq. (\ref{wkbkuniform}) and
is an average of the Fourier basis over some bin centered at $k$,
with width $\Delta k$.

It can be shown that the variance of such a bin-averaged power
estimate is given by (Appendix B)
\begin{mathletters}
\label{Ek}
\begin{eqnarray}
\label{Vkk}
&& V (k) \equiv \langle [\hat P_2 (k) - P (k)]^2 \rangle 
= {{\cal N}^3 \over {\cal L}^2} \sum_\alpha  
[{w}^{\alpha\alpha} (k)]^2  E^{\alpha} (k) \\ 
\label{Ekdef}
&& E^{\alpha} (k) \equiv {2\over
{n_{\bar k}}} \left[P (k) +
{{\cal L}\over {\cal N}} {q^\alpha \bar N_{\rm 
Q}^\alpha + V_{\rm 
B}^\alpha \over (\bar N_{\rm Q}^\alpha)^2} \right]^2 + 
{1\over \cal L} \langle T \rangle_{kk} \\ \nonumber && \quad \quad + 4
\langle B
\rangle_{kk} {q^\alpha \over \bar N_{\rm Q}^\alpha} {1 \over
{\cal N}} + 
2 \langle P \rangle_{kk} \left[{q^\alpha \over \bar
N_{\rm Q}^\alpha}\right]^2 { {\cal L} \over {\cal N}^2 } \\
\nonumber
&& \quad \quad + 4 P(k) {{q'}^\alpha \over (\bar N_{\rm Q}^\alpha)^2}
{\cal L \over \cal N} + {{\cal L}^2 \over {\cal N}^3}
{{q''}^\alpha \over ({\bar N_{\rm Q}^\alpha})^3}
\end{eqnarray}
\end{mathletters}
assuming the $k$ of interest satisfies $k \gg 1/{\cal L}$ and
that the width of the bin $\Delta k$ also satisfies $\Delta k \gg
1/{\cal L}$, where ${\cal L}$ is the length of the spectrum.  This is
sometimes referred to as the classical 
limit in the case of galaxy power spectrum measurement
(\citenp{fkp94,hamilton97a}). We will not consider larger scales
here, because measurements on such scales are likely dominated
by systematic rather than random errors.

The symbol $n_k$ denotes the number of modes within the k bin of
interest, ${\cal N}$ is the number of pixels in the length ${\cal L}$;
$q^\alpha$, $\bar N_{\rm Q}^\alpha$ and $V_{\rm B}^\alpha$ are as
defined in eq. (\ref{bkuniform}). The quantities $q'^\alpha$
and $q''^\alpha$ are analogous to $q^\alpha$:
\begin{equation}
q'^\alpha \equiv \sum_i (W^{i\alpha})^3 g_{\rm ps}^{i\alpha} g_{\rm
b}^\alpha \, \, , \, \, q''^\alpha \equiv
\sum_i (W^{i\alpha})^4  g_{\rm ps}^{i\alpha} g_{\rm
b}^\alpha 
\label{qpqpp}
\end{equation}
The symbols
$\langle T \rangle_{kk}$, $\langle B \rangle_{kk}$ and $\langle P
\rangle_{kk}$ represent
the shell-averaged trispectrum, bispectrum and power spectrum
respectively ($\langle \, \rangle$ here is to be distinguished from
ensemble average discussed in \S \ref{notation}): 
\begin{mathletters}
\label{PBT}
\begin{eqnarray}
\label{Tave}
\langle T \rangle_{k_1 k_2} \equiv && {1\over n_{k_1}
n_{k_2}} \sum_{k_1'} \sum_{k_2'} T (-k_1',k_1',-k_2',k_2') \\
\label{Bave}
\langle B \rangle_{k_1 k_2} \equiv && {1\over n_{k_1}
n_{k_2}} \sum_{k_1'} \sum_{k_2'} B (-k_1'-k_2',k_1',k_2') \\
\label{Pave}
\langle P \rangle_{k_1 k_2} \equiv && {1\over n_{k_1}
n_{k_2}} \sum_{k_1'} \sum_{k_2'} P (k_1' + k_2')
\end{eqnarray}
\end{mathletters}
where the sum over $k_1'$ extends over modes within the bin centered
at $k_1$, and similarly for $k_2'$. 
The trispectrum $T$ and bispectrum $B$ are Fourier transforms of
the four and three point correlation functions, defined in an analogous 
manner
to eq. (\ref{Pxidef}). 

The variance as given in eq. (\ref{Ek}) contains contributions
from both cosmic fluctuations and discrete fluctuations (see \S \ref{notation}).
The terms such as $P(k)^2$ and $\langle T \rangle_{kk}$ 
arise because of intrinsic fluctuations of the cosmic
signal from one part of the universe to another --
these terms are present even if one has data with
arbitrarily high signal-to-noise. The terms containing
$\bar N_{\rm Q}^\alpha$ arise because of discrete fluctuations
-- these we will loosely referred to as shot-noise.

As we have emphasized in \S \ref{describe} \&
\ref{estimator}, the shot noise
contributions to the random error are not exactly Poisson-distributed. 
The shot noise contributions (ignoring cosmic sample fluctuations)
in eq. (\ref{Ekdef}) would all be
simply $1/\bar N_{\rm Q}^\alpha$ if $\hat N_{\rm Q}^\alpha$ were 
strictly a
Poisson variable. We have
additional fluctuations in $\hat N_{\rm Q}$ due to the background (sky
and readout), and
also due to non-unity weights used in reducing the data (eq.
[\ref{Ncount}]; see also end of \S \ref{describe}). 

Given eq. (\ref{Ek}), it is simple to derive a weighting ${\bar
w}^{\alpha\beta} (k)$ that minimizes the variance $V (k)$,
subject to the constraint that the effective window
($G$ as defined in eq. [\ref{Gwindow}]) is properly normalized.
This is most simply derived by minimizing the following Lagrangian:
\begin{equation}
L (k) = V (k) - \lambda ( \int G(k,k') {dk'\over 2 \pi} - 1)
\label{Lagrangian}
\end{equation}
where $\lambda$ is a Lagrange multipler. Differentiating
the above respect to ${\bar
w}^{\alpha\beta} (k)$, and setting the result to zero,
we obtain:
\begin{equation}
{\bar w}^{\alpha\beta} (k) = [E^{\alpha} (k) E^{\beta} (k)]^{-{1\over
2}} / M (k) \, \, \, \, , \, \, \, \, M (k) \equiv  
\sum_{\mu} [E^{\mu} (k)]^{-{1}} {{\cal N}/{\cal L}}
\label{minw}
\end{equation}
The corresponding shot-noise subtraction, instead of eq.
(\ref{bkuniform}), would then be
\begin{equation}
b (k) = \sum_\alpha {\bar w}^{\alpha\alpha} (k) {q^\alpha \bar N_{\rm
Q}^\alpha +  
V_{\rm B}^\alpha \over (\bar N_{\rm Q}^\alpha)^2}
\label{minb}
\end{equation}
where $q^\alpha$, $N_Q^\alpha$ and $V_B^\alpha$ are as defined in eq.
(\ref{bkuniform}).

In summary,
the minimum variance estimator of the power spectrum is
%given by eq. (\ref{hatP}), (\ref{barw}), (\ref{minw}) \& (\ref{minb}),
%together with trend removal as explained in \S \ref{continuum} i.e.
\begin{equation}
\hat P_5 (k) = \sum_{\alpha,\beta} R^{\alpha\beta} (k) 
[E^\alpha (k)^{1\over 2}
\hat \delta_f^\alpha] [E^\beta (k)^{1\over 2}
\hat \delta_f^\beta] / M(k) - b (k)
\label{P5}
\end{equation}
where $b (k)$ is given by eq. (\ref{minb}), $E^\alpha (k)$ and
$M (k)$ are given in eq. (\ref{Ek}) and (\ref{minw}),
and $R^{\alpha\beta} (k)$ is as in eq. (\ref{wkuniform}).
%Note that an important problem which the above estimator does not
%address is the cross-correlation between different wave bands, which
%would be present if $\hat \delta_f$ were significantly non-Gaussian
%(\citenp{mw98,szh99,hamitlon99}).

The minimum variance estimator can be understood simply as follows:
$\hat \delta_f^\alpha$ at each pixel is weighed by $1/\sqrt {E^\alpha (k)}$ 
before the array is Fourier transformed, squared and grouped to form band
power estimates. Note that the above estimator reduces to the one with uniform weighting (eq. [\ref{wkbkuniform}]) if $E^\alpha (k)$ were 
independent of
$\alpha$, e.g. when sample/cosmic variance is significantly larger
than shot-noise ($P (k) \gg [{\cal L}/{\cal N}] [q^\alpha \bar N_{\rm
Q}^\alpha +V_{\rm B}^\alpha] / [\bar N_{\rm Q}^\alpha]^2$). 
It is important to note that 
the weighting as a function of $\alpha$ is determined
by $\bar N_{\rm Q}^\alpha$ rather than, say $\hat N_{\rm Q}^\alpha$.
Down-weighing pixels with a lot of absorption
(hence relatively low $\hat N_{\rm Q}^\alpha$)
would be a wrong thing to do, since the fluctuations in absorption
is the signal that we are after. The proper procedure
is to down-weigh pixels with an overall lower mean count 
$\bar N_{\rm Q}^\alpha$.

Unfortunately, the minimum variance weighting given above is difficult
to implement because one needs to specify simultaneously $P$, $B$ and
$T$, in addition to the level of shot-noise.
A common simplification is to use the Gaussian approximation in which
$E^\alpha (k)$ is approximated as:
\begin{equation}
E^{\alpha} (k) \sim  {2\over
{n_{\bar k}}} \left[P (k) +
{{\cal L}\over {\cal N}} {q^\alpha \bar N_{\rm 
Q}^\alpha + V_{\rm 
B}^\alpha \over (\bar N_{\rm Q}^\alpha)^2} \right]^2 
\label{EkGauss}
\end{equation}
(see e.g. \citenp{hamilton97a}). Note that in addition to ignoring the
bispectrum and trispectrum terms, the above also ignores
certain power spectrum terms
which are mixed with shot-noise -- the last three
terms in eq. (\ref{Ek}), which
is equivalent to assuming that either the shot-noise or the
correlation is sufficiently weak.
With the above approximation, one can
start with some initial $P$ and use the minimum variance weighting scheme
to get a first measurement of $P$, and iterate subsequently 
(\citenp{bjk97}). 
Analogous (Gaussian) power spectrum estimators
for galaxy-surveys and microwave background experiments
have been widely discussed in the literature
(e.g. \citenp{fkp94,vs96,tth97,hamilton97a,tegmark98,bjk97,seljak97}).

We will not attempt to address here the important question of how 
significant the non-Gaussian contributions
are. A proper treatment will involve the analysis of
a large number of simulations or a large data-set, which
we hope to present in a future paper.
It suffices to say that 
the very nonlinear mapping from the density field to $e^{-\tau}$ will
likely introduce some degree of non-Gaussianity, even if the initial 
density field is Gaussian.

The use of observed data to study this issue is particularly
interesting, and deserves some more comments. 
In principle, since different QSO sightlines typically sample
independent regions of the universe, one can estimate the variance
of the transmission power spectrum,
and hence infer the importance of the non-Gaussian
contributions, using directly the fluctuations in
power spectrum estimates from one sightline to another.
However, one should keep in mind that shot-noise also
contributes to the variance. Since different lines of sight
generally have different S/N, the sightline-to-sightline
fluctuations in power spectrum estimates should be interpreted
with care. In a dataset of several quasars, it is possible
that the quasar-to-quasar fluctuations are dominated by
a few with low S/N, and their mean-square would give
an overestimate of the true power spectrum variance.

%based on OLD highKeckNM.C.FKP.plot
%highKeckNM.C.FKP2.plot
%\begin{figure}[htb]
%\centerline{\psfig{figure=highKeckNM.C.FKP.ps,height=7.0in}}
%\caption{Power spectrum estimation using uniform weighting (bottom
%panel; eq. [\ref{wkbkuniform}]) versus minimum variance weighting (top
%panel; eq. [\ref{minw}]). The simulated QSO spectrum consists of 12
%segments, half of which have comparable S/N to Fig. \ref{highKeckNM.C}
%and half of which have $\sim 20$ times lower S/N.} 
%\label{highKeckNM.C.FKP}
%\end{figure}

\begin{figure}[htb]
\centerline{\psfig{figure=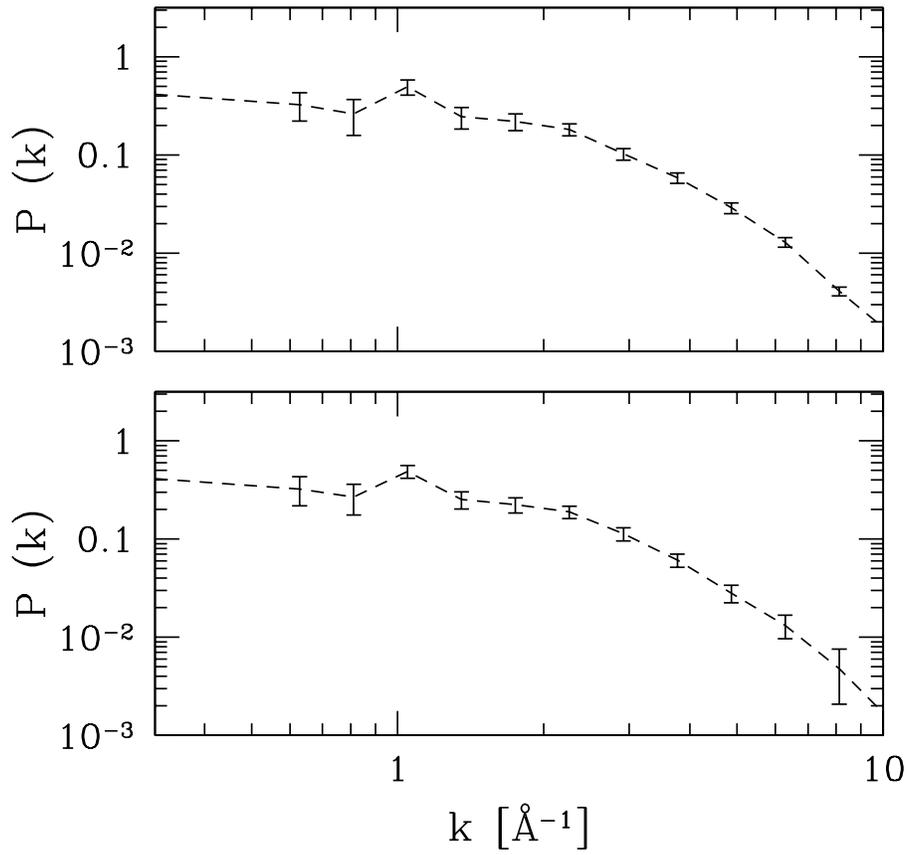,height=5.0in}}
\caption{Power spectrum estimation using uniform weighting (bottom
panel; eq. [\ref{wkbkuniform}]) versus minimum variance weighting (top
panel; eq. [\ref{minw}]). The simulated QSO spectrum consists of 12
segments, half of which have comparable S/N to Fig. \ref{highKeckNM.C}
and half of which have $\sim 20$ times lower S/N.} 
\label{highKeckNM.C.FKP}
\end{figure}

We show in Fig. \ref{highKeckNM.C.FKP} an example in which
the data consist of 6 high quality spectra (similar
to Fig. \ref{highKeckNM.C}) and 6 others with S/N about 20
times smaller. The bottom panel shows the power spectrum estimated with
uniform weighting (eq. [\ref{wkbkuniform}]) while the top panel represents
the power spectrum estimated with minimum variance weighting using
the Gaussian approximation. The (1 $\sigma$) error-bars are theoretical
-- they are 
estimated using eq. (\ref{Vkk}) and (\ref{EkGauss}).
%\footnote{It is important that the error-bars are estimated with a
%smooth version of the band-power estimates to avoid overestimation or
%underestimation of error-bars due to fluctuations in individual bands
%(see e.g. \citenp{seljak97,bjk97}).}
This illustrates how our weighting scheme can reduce the 
error bars at high $k$'s where shot-noise is important.
Absent information on the non-Gaussian nature of
the power spectrum variance, we advocate the Gaussian
weighting scheme (eq. [\ref{EkGauss}]) as a rational way
to combine data with different levels of S/N to reduce
the variance, even though it does not necessarily achieve
minimum variance.

In combining the different spectra
with different S/N, we have weighed the power spectrum
estimate of each line of sight by its inverse variance, which is
an obvious generalization of the minimum
variance weighting introduced above. For instance, suppose we have
two separate lines of sight $A$ and $B$, we could combine the two power
spectrum estimate $\hat P_2^A$ and $\hat P_2^B$ in the following way,
assuming the two lines of sight are independent:
\begin{equation}
\hat P_2 (k) = \left[ {\hat P_2^A (k) \over V_A (k)} + {\hat P_2^B (k)
\over V_B (k)} \right] / \left[ {1\over V_A (k)} + {1\over V_B (k)}
\right] 
\label{weightLOS}
\end{equation}
where $V_A (k)$ and $V_B (k)$ are estimated with the same $P$ but
could have different levels of shot-noise. The noisier quasar spectrum
is naturally down-weighed.

Lastly, we should emphasize the above discussion does not address
the issue of cross-variance between power spectrum estimates
at two different wave bands, which is introduced by the non-Gaussian
terms (\citenp{mw98,szh99}). Hamilton \cite*{hamilton99}
introduced a scheme which simultaneously diagonalizes
the covariance and minimizes it. However, it makes specific assumptions
about the form of the trispectrum and bispectrum, the validity
of which for the forest remains to be checked.

\section{Discussion}
\label{conclude}

Our recipe for measuring the transmission power spectrum is
summarized here.

\begin{itemize}

\item Given an array of reduced quasar counts 
$\hat N_{\rm Q}^\alpha$, identified metal
lines should be removed, especially the strong ones ($\tau > 1$).
Small gaps in the spectrum (e.g. due to cosmic-ray-hits removal)
can be (linearly) interpolated over, while
large gaps should be avoided (\S \ref{gaps}). 

\item The mean quasar counts ($\bar N_{\rm Q}^\alpha$) 
is estimated using eq. (\ref{Nmeanest}). The mean-basis (the
functional form of the mean-trend)
should be chosen to be as smooth as possible -- underfitting is better
than overfitting (see Test 1 and 2 of \S \ref{continuum}).
In practice it appears that a flat mean suffices for short spectra
($\sim 50 \AA$), while polynomials up to the third order
can be used for longer spectra ($\sim 500 \AA$).
One can gain an idea of what a reasonable basis is using 
the red side of Ly$\alpha$ or low redshift QSO spectra
where the continuum can be seen more
clearly.

\item Define the trend-removed and normalized 
fluctuation $\hat \delta_f^\alpha$
according to eq. (\ref{hatdeltaf}), and the power spectrum is
estimated using the quadratic estimator in eq. (\ref{hatPP}).
Different weightings are possible, the simplest that we recommend is
given in eq. (\ref{wkbkuniform}).
A more sophisticated weighting scheme which can reduce
the random error is given by
eq. (\ref{minw}), (\ref{minb}), (\ref{P5}) and (\ref{EkGauss}).
If one is interested in the real-space correlation function instead,
the recommended weighting is eq. (\ref{wu}) -- this gives
a smaller variance compared to other estimators commonly used
in the literature. We emphasize that the shot-noise-bias
($b (k)$ in eq. [\ref{bkuniform}], or in eq. [\ref{minb}])
should be subtracted correctly, especially for noisy data.

\item If better control over systematic errors on large scales 
($ \approxgt \, 30 \AA $) 
introduced by the unknown continuum is desired, the techniques
outlined in \S \ref{beyond} can be used. The corresponding
estimator is given in eq. (\ref{P3}), which requires
an estimate of the continuum power spectrum $P_{\rm C}$
(eq. [\ref{expectP2b}]) and 
an additional correction factor $\epsilon_{\rm G}$
(eq. [\ref{epsilonG}]). This procedure requires the 
identification of an appropriate set of continua (see 
discussions in \S \ref{beyond}). Even if one is not
interested in the power spectrum on large scales, we recommend
this procedure as a consistency check that the spurious
power introduced by the continuum is negligible on the scales
of interest.

\end{itemize}

What implications does the above have for one's observing strategy?
To discuss this question, we need to take a closer look at
the issue of shot-noise.
The shot-noise enters in two different places in the
above discussion. 
First, it contributes to the variance (random error)
of the power spectrum estimate (eq. [\ref{Ek}]).
Second, it appears as a bias in the power
spectrum estimate that we have to subtract off
(e.g. $b(k)$ in eq. [\ref{hatPP}] \& [\ref{wkbkuniform}]).

In the literature on power spectrum measurement,
shot-noise subtraction has been largely ignored (e.g. 
\citenp{croft98}; see \citenp{mcdonald99} for an alternative
approach where shot-noise is simulated rather than subtracted).
Let us estimate how important it is.
The expression in the simplest case of uniform weighting
is given in eq. (\ref{bkuniform}) (see eq. [\ref{minb}] for
more complicated weightings), which can be rewritten as
\begin{equation}
b (k) = {\Delta u \over {\cal N}} \sum_\alpha
{\sum_i (W^{i\alpha})^2 [ g_{\rm ps}^{i\alpha} g_{\rm
b}^\alpha \bar N_Q^\alpha + V_B^{i\alpha} ]
\over (\bar N_Q^\alpha)^2}
\label{brewrite}
\end{equation}
where $\Delta u$ is the size of a pixel, ${\cal N}$ is the total
number of pixels, and the rest of the symbols are as defined in
\S \ref{describe}: 
$i$ is the pixel-label in the spatial direction
and $\alpha$ in the spectral direction, $\bar N_{\rm Q}^\alpha$
is the mean reduced quasar count, $V_{\rm B}^{i\alpha}$ is the background
variance, $W^{i\alpha}$ is a weighting, and $g_{\rm ps}^{i\alpha}$ and
$g_{\rm b}^\alpha$ are the point-spread function and blaze function
respectively (eq. [\ref{W1}] \& [\ref{ggNQ}]). 
An important observation is that the
numerator within the summation is closely related to the
variance array which is often given along with a spectrum
(eq. [\ref{deltapN}])\footnote{See paragraph after eq. (\ref{deltapN}) and \S \ref{notation}
on the distinction between $\tilde N_{\rm Q}^{i\alpha}$ and
$\hat N_{\rm Q}^{i\alpha}$.}:
\begin{equation}
{\, \rm var \,} (\alpha)= 
{\sum_i (W^{i\alpha})^2 [ \hat N_{\rm Q}^{i\alpha} + V_B^{i\alpha} ]}
\label{varalpha}
\end{equation}
The quantity $\hat N_{\rm Q}^{i\alpha}$ is of course
different from $g_{\rm ps}^{i\alpha} g_{\rm
b}^\alpha \bar N_Q^\alpha$ which we need to estimate the shot-noise,
but since we are in practice interested in an average over all
pixels, it turns out the following estimate of the shot-noise
is accurate to within a percent for all cases we have tested:
\begin{equation}
b(k) \sim {\Delta u \over {\cal N}} \sum_\alpha {{\, \rm var \,} (\alpha)
\over (\bar N_Q^\alpha)^2}
\label{bkapprox}
\end{equation}
Without the above approximation, an exact estimate of the shot-noise
would require the knowledge of $\hat N_Q^\alpha$, 
$\sum_i (W^{i\alpha})^2 g_{\rm ps}^{i\alpha} g_{\rm b}^\alpha$
and $\sum_i (W^{i\alpha})^2 V_B^{i\alpha}$.

Eq. (\ref{bkapprox}) provides a useful means of estimating
the shot-noise (see Appendix A on shot-noise estimation 
under more complicated circumstances i.e. with
non-trivial rebinning or weighting). One can simplify further
by making a crude approximation in 
relating $b (k)$ to the typical signal-to-noise
ratio of the data (which is often quoted at the continuum) through
\begin{equation}
b (k) \sim (\Delta u / \bar f) (N/S)^2 
\label{bkSN}
\end{equation}
which can be justified if one ignores the part of the variance
due to the sky and readout. We find that this simple rule
of thumb generally provides
an underestimate of the shot-noise (particularly at low
$S/N$ where the background counts become important), 
but is accurate to within
about a factor of $2$.

\begin{figure}[htb]
\centerline{\psfig{figure=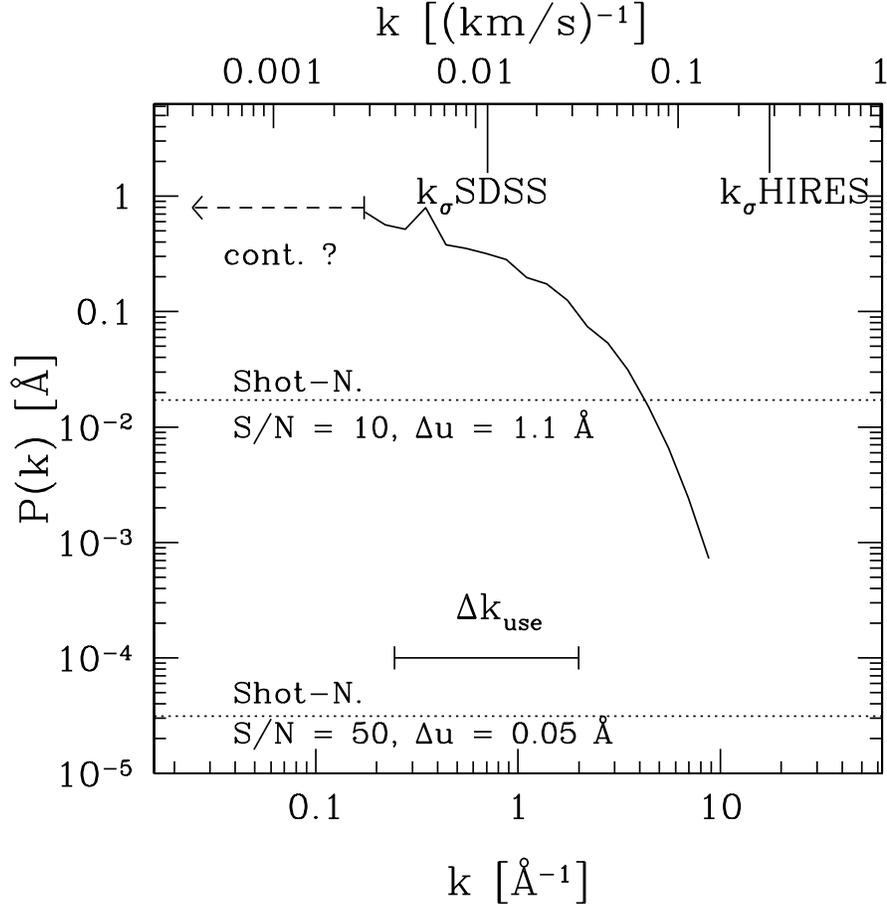,height=5.0in}}
\caption{
The solid line is the mean observed transmission power
spectrum at $z = 3$ from McDonald et al. (1999).
Its continuation as a dashed line at small $k$'s indicate
the scales at which continuum might introduce significant 
uncertainties, and so the power is not shown. 
The two dotted lines at top and bottom shows the level of
shot-noise at two extremes. The top one resembles the quality
expected for SDSS spectra while the bottom resembles HIRES
Keck spectra of bright quasars. The interval of $\Delta k_{\rm use}$ 
at the bottom indicates the range of scales that are currently
used to recover the linear mass power spectrum. The long
tickmarks at the top indicates the resolution of SDSS and typical Keck
HIRES spectra. Different units of
$k$ are related by $ [k / {\, \rm (km/s)^{-1}}] = [(1+z) / 4] [k /
{\AA}^{-1}] / 61.67 = [(1+z) / \sqrt{\Omega_m (1+z)^3 + \Omega_k 
(1+z)^2 + \Omega_\Lambda}]
[k / {\, \rm h Mpc^{-1}}] / 100$.
}
\label{demonstrate}
\end{figure}

Fig. \ref{demonstrate} summarizes some useful information for devising
an observing strategy, based on our estimate of the shot-noise
in eq. (\ref{bkSN}) above. The solid line shows the mean observed 
transmission power spectrum at $z = 3$ taken from McDonald et al. 
\cite*{mcdonald99}.\footnote{We divide the
un-normalized power spectrum (eq. [\ref{Pxidef0}])
of McDonald et al. by the square of their measured
mean transmission to obtain the normalized power spectrum given
in Fig. \ref{demonstrate}. See tests 4 and 5 in \S \ref{continuum}
on the bias of the un-normalized power spectrum.}
The two horizontal dotted lines show
the level of shot-noise expected for the 2 extremes of
the kinds of observations we are likely to encounter --
the bottom corresponds to very high signal to noise (S/N) observations
with HIRES quality resolution (e.g. \citenp{hu95,kirkman97,rauch97}),
while the dotted line on the top corresponds to low S/N
observations expected for a large number of quasars in the 
Sloan Digital Sky Survey (SDSS). We emphasize that the shot-noise
level does not depend on the resolution per se, but on the 
pixel size for a given S/N. 
SDSS is expected to produce $\sim 1000$ QSO spectra at 
S/N $= 20$ per pixel (QSO's at 
$z > 2.7$, where the redshift limit is determined by the
blue limit of the spectrograph, $3800 \AA$),
$\sim 10000$ at S/N $= 15$ and $\sim 30000$ at S/N $=7$,
corresponding to $i'$ magnitude-cuts at $18$, $19$ and $20$ respectively
(\citenp{fukugita96,fan99}).
The pixel size of SDSS is quite uniform in velocity $70 {\, \rm km/s}$,
which is equivalent to $1.13 \AA$ at $4864 \AA$ (Ly$\alpha$ at $z = 3$).

Clearly, the importance of shot-noise depends on the scales
at which one is interested in measuring the power spectrum.
A few interesting scales are shown in Fig. \ref{demonstrate}.
First, instrumental resolution imposes a high $k$ limit beyond
which one cannot reliably measure the transmission power spectrum.
The resolution window is often characterized by a Gaussian with
a given FWHM. The effect of such a resolution window on
the power spectrum can be represented by $P (k) \rightarrow P (k)
e^{-k^2/k_\sigma^2}$ where $k_\sigma = \sqrt{2/{\rm ln} 2} / 
{\, \rm FWHM
\, }
\sim 1.7 / {\, \rm FWHM \, }$. Two representative $k_\sigma$'s are
shown as long tickmarks at the top. Note that even at $k = k_\sigma$,
the resolution window reduces the power by $63 \%$ and so
has a non-negligible effect. The Sloan FWHM is about $2.1$ pixels
i.e. $147 {\, \rm km/s}$, or $2.4 \AA$ at $4864 \AA$.

On the other hand, 
the range of scales that is currently being used to infer
the {\it mass} power spectrum is indicated by the interval
near the bottom $\Delta k_{\rm use}$. The high $k$ limit
is set by the scales at which the shape of the power spectrum is 
preserved in the transformation from mass to transmission
(i.e. linear biasing e.g. \citenp{croft98}). We can see that
for high quality Keck spectra, information from a whole decade
of measurable scales is unused for the recovery
of the mass power spectrum -- it would be very useful
to push the current analysis techniques to these scales,
since power on these scales is of particular interest in
constraining e.g. neutrino properties (\citenp{hui97a,croft99}).
Such an effort would require disentangling the effects of
peculiar velocities and thermal broadening,
however. At the other end, the low $k$ limit of currently usable
scales is set by the scales at which the continuum fluctuates.
This is indicated by the dashed line at the top, where the
transmission power spectrum is unknown.

From the above discussion, we can distill a few tips for 
observing/analysis.

\begin{itemize}

\item To ensure that shot-noise is subdominant, one 
might want to achieve $S/N \, \approxgt \,$
$\sqrt{ 10 \Delta u / \bar f / P(k_{\rm int.})}$
where $k_{\rm int.}$ is the scale of interest, and $\bar f$ is
the mean transmission. The factor of $10$ is somewhat 
arbitrary -- this will ensure the shot-noise contribution to
the power spectrum variance is no more than about $20 \%$
(under the Gaussian approximation; see eq. [\ref{EkGauss}]),
or the $1 \sigma$ error-bar on the power spectrum would
only be increased by $10 \%$ due to the contribution from shot-noise.
An important question is what $k_{\rm int.}$ should be --
that depends on at what scales one can usefully extract 
cosmologically interesting information. Current literature
mainly focused on $k_{\rm int.} \, \approxlt \, 2 \AA^{-1}$, where 
$P \sim 0.06 \AA$, therefore $S/N \sim 20 \sqrt{\Delta u/1 \AA}$
would be sufficient. Since $P$ rises with scale, shot-noise
would be even less important at smaller $k$'s.
Note that with very small $\Delta u$ such as $\sim 0.05 \AA$,
$S/N$ as low as $4 - 5$ is acceptable.
To give some examples, a $S/N$ of 8 per $0.05 \AA$ can be achieved
with an hour of exposure using Keck/HIRES for a $V = 19$ quasar;
on the other hand, a $S/N$ of about 15 per $1.1 \AA$ 
is expected with just slightly under an hour of exposure
using the SDSS spectrograph for a $i' = 19$ quasar.

\item A corollary of focusing on only $k \, \approxlt \, 2 \AA^{-1}
$ is that observations with $k_\sigma \,
\approxgt \, 3 \times 2 \AA^{-1}$
or a resolution FWHM of $0.3 \AA$ or $R \sim 16000$ at
$z = 3$ are acceptable.
The factor of $3$ above (i.e. in $3 \times 2 \AA^{-1}$)
is somewhat arbitrary --
it ensures that at $k = 2 \AA^{-1}$, the resolution window
does not reduce the power by more than $10 \%$. If the resolution
window is known accurately, or if one is willing to
sacrifice information on the small scales close
to $k \sim 2 \AA^{-1}$, one could in principle consider
lower resolutions.
We would like to emphasize, however, that
in principle, the modes at $k > 2 \AA^{-1}$ could still contain
very interesting cosmological information, even though the
current attempts at recovering the mass power spectrum ignored
them.

\item If shot-noise is subdominant compared to the power
spectrum, the only other limiting factor to the size
of the random error is the total size of one's
sample or the number of sightlines in it.
Assuming all sightlines have similar coverage with length
${\cal L}$, then the fractional error of a single k-mode
(i.e. in a k-bin of $2 \pi / {\cal L}$) is given by 
$\delta P / P = C/\sqrt{N_{\rm sight}}$ where 
$N_{\rm sight}$ is the number of sightlines assuming they
are independent, and 
$C = 1$ under the Gaussian approximation (eq. [\ref{EkGauss}]), 
and a little larger than unity under more general 
circumstance (see e.g. \citenp{mw98,szh99}). 

\item How should one distribute one's observing time among quasar
targets to minimize the random error on the transmission
power spectrum? There are many possible versions of this problem.
We will discuss two, giving an explicit solution for the first,
and only general expressions for the second.
In the simplest case in which all the candidate quasar targets have
similar magnitudes, given a finite amount of observing time,
one can deduce the optimal total number of quasars one
should target by
\begin{equation}
{\rm minimizing} \quad {N_{\rm tot.}}^{-1} [ P(k_{\rm int.})
+ {\Delta u \over \bar f} {A \over t} ]^2 \quad \quad \quad 
{\rm subject \, \, \, to} \quad 
N_{\rm tot.} t = T_{\rm tot.}
\label{Ntot}
\end{equation}
where $N_{\rm tot.}$ is the total number of quasars targeted,
$k_{\rm int.}$ is the scale of interest,
$T_{\rm tot.}$ is the total amount of observing time one has,
$t$ is the amount of time one spends on each quasar, and
$1/A$ is equal to $(S/N)^2$ reached per unit exposure time.
The above assumes eq. (\ref{EkGauss}) and that the sightlines are
independent. The solution is easy to deduce:
$N_{\rm tot.} = [P(k_{\rm int.}) \bar f / \Delta u] [T_{\rm tot.} / A]$,
or $t = A \Delta u  / \bar f / P(k_{\rm int.})$.
A typical value for $1/A$ is $1/A \sim 1200
{\, \rm hour} {\, \rm per \, \, \AA}\, \, \times 
10^{(19 - mag.)/2.5} \times [{\, \rm aperture} / 100 {\, \rm m.}^2] 
\times f_{\rm throughput}$, where $f_{\rm throughput}$ is
about unity for Keck/HIRES, and $\sim 2.5$ for the SDSS.
Using again $k_{\rm int.} \sim 2 \AA^{-1}$, for a $19$th magnitude
quasar, with an aperture of $ 6.25 {\, \rm m.}^2$ and assuming
$\Delta u = 1 \AA$ and $f_{\rm throughput} = 3$, the exposure
time is $t = 14$ minutes only! 
The above prescription, however, only allows for just enough exposure
time to reduce the shot-noise to a level comparable to the
cosmic/sampling variance (i.e. $P(k_{\rm int.}) \sim (\Delta u / \bar f)
(N/S)^2$) -- the sole aim is to maximize the number of quasars observed
within a given length of time to beat down the sampling variance. 
The prescription would certainly be different
if one has, for instance, a finite number of quasar targets, or
if one has other purposes in mind -- such as measuring the
mean decrement, etc (see earlier prescription for making shot-noise
subdominant, equivalent to multiplying $t$ by about a factor of
$10$). 
A more general version of the above problem deals with a case
where the quasars span a range of magnitudes
i.e. $A$ is no longer the same number for each quasar. A simple
ansatz is to assume $t = \alpha 
A \Delta u / \bar f / P(k_{\rm int.})$,
in other words, spending more time for fainter quasars because
it takes longer to beat down the shot-noise, except that
we have a normalizing factor $\alpha$ which enforces the
constraint of total observing time: $\alpha =
T_{\rm tot.} [\bar f P(k_{\rm int.})/\Delta u] /
\int_{A_{\rm min}}^A A' n(A') dA'$
where $n(A) dA$ is
the number of quasars with $A$ falling in the given range, and
$A_{\rm min}$ corresponds to the brightest quasar in one's sample.
Then, we can determine how many quasars one should include, 
starting from the brightest one, or
how faint one should go by minimizing
$[\int_{A_{\rm min}}^A n(A') dA']^{-1}
P(k_{\rm int.})^2 (1 + 1/\alpha)^2$ with respect to $A$.

\end{itemize}

{\it The following is particularly relevant for SDSS or comparable
observations.}

\begin{itemize}

\item In addition to contributing to the power spectrum variance,
shot-noise also contributes a bias which has to be subtracted off
(see e.g. Fig. \ref{LickNM_cont3L}).
This is quite important for SDSS because, with 
$> 10^4$ sightlines,
the survey has the capability of reducing the fractional error
of the power spectrum to $< 1 \%$ per mode. Therefore, 
a bias of $\sim 3 - 100\%$, depending on the scale of interest
(as indicated by the top
dotted line in Fig. \ref{demonstrate}), is not acceptable and
should be subtracted
off. We note that analyses so far in the literature (e.g. \citenp{croft98b,mcdonald99}) focused
on higher quality data where $S/N \sim 30$, with
$\Delta u$ ranging from about $0.04 \AA$ to $1 \AA$, and so
according to eq. (\ref{bkSN}) and Fig. \ref{demonstrate}, the
shot-noise bias was about $1 \%$ of the power or smaller
and therefore could be ignored, although a more careful check should
be performed for some datasets with lower $S/N$.

\item The low resolution of SDSS spectra implies that it would
be difficult to obtain useful information on scales $k \gg k_\sigma
\sim 0.01 {\, \rm (km/s)^{-1}}$ or $0.7 \AA^{-1}$. 
On larger scales or smaller $k$'s,
two problems have to be reckoned with. 
For $k \sim 0.3 k_\sigma - k_\sigma$,
the resolution window suppresses the power by $10 \%$ or more --
therefore, one needs to have an accurate measure of the resolution
window to recover the true transmission power spectrum. 
\footnote{We thank Rupert Croft for useful discussions on this
point.}
This can be achieved by using narrow metal lines or arc lines. There
are relatively few sky lines in the relevant part of the spectrum.

\item For scales $k < 0.004 {\, \rm (km/s)^{-1}}$ or
$0.2 \AA^{-1}$, the effect of the continuum has to be properly
taken into account, and the method of \S \ref{beyond} can
be useful here. From Fig. \ref{demonstrate}, it is clear that 
the range of scales accessible to SDSS would be quite limited
unless a correction for continuum contamination is applied.

\end{itemize}

It is worth pointing out that while the above quoted numbers are
all based on $z = 3$, we do not expect them to change significantly
for $z =2$ or $z = 4$. This is in part because of the slow evolution
of the transmission power spectrum -- the growth of the mass
spectrum with time is partially
compensated by the lowering of the mean decrement
(\citenp{mcdonald99}). 

At least three issues remain to be explored in future work. 
As we have emphasized, the concept of correcting for continuum
contamination in the transmission power on large scales as laid
out in \S \ref{beyond} is an interesting one, but requires
more testing. An important check is the measurement of 
continuum power spectrum as a function of redshift on both
sides of Ly$\alpha$ emission.
Second, counts-in-cells analysis (i.e. measuring moments of the
one-point probability distribution function), just
like power spectrum analysis, requires shot-noise subtraction,
and since typically one considers cells at the limit of
resolution, shot-noise is likely non-negligible
except for high S/N spectra. Counts-in-cells analysis
provides a very interesting way to test the gravitational 
instability paradigm (\citenp{gc98,hui98b,adi99}), and should be done 
with care. Useful expressions will be presented elsewhere
(\citenp{hui00}). Lastly, as is hopefully clear by now, 
the spirit of the methods presented in this paper is to avoid
continuum-fitting and replace it with trend-removal. 
We have demonstrated that
this is possible for
measuring the transmission power spectrum. However, other quantities
of interest such as the mean decrement requires an estimate of
the continuum, by definition. Furthermore, to theoretically
interpret the transmission power spectrum in terms of the
mass fluctuation, current methods require the measurement of
the mean decrement to fix a free parameter in one's cosmological
model, which is a combination of the ionizing background,
the mean baryon density and the mean temperature. Therefore,
in a sense, the technique of trend-removal only goes half-way in 
solving the problem of continuum-fitting. Although we still
recommend our method over continuum-fitting because the transmission
power spectrum is an unambiguous quantity which can and should
be determined as accurately as possible (not to mention
the fact that continuum-fitting is difficult with low S/N or
low resolution, or at high $z$), there is clearly a need for
an alternative method to bridge the 
gap between the measured transmission
power and the theoretically interesting 
mass power. This will be explored in future 
publications (\citenp{hb00,zsh00}).

We thank for useful discussions Len Cowie, David Kirkman, Patrick Petitjean, 
Michael Rauch, Wal Sargent, Don Schneider, 
David Weinberg, Don York
and the participants of
the 1999 Haifa workshop on the intergalactic medium and large
scale structure. 
We also thank Nick Gnedin for supplying
a simulation. Special thanks are due to
Matias Zaldarriaga for pointing out the importance
of aliasing and for many interesting discussions. 
This work was supported in part by
the DOE and the NASA grant NAG 5-7092 at Fermilab, and
by the NSF grant PHY-9513835. LH thanks the IAS for the
Taplin Fellowship.

\section*{Appendix A}
\label{app1}

Our main aim here is to derive eq. (\ref{Gwindow}) for the estimator
$\hat P_2 (k)$ which is
given by eq. (\ref{hatP}) and eq. (\ref{wkbkuniform}), with an eye
towards generalization to $W^{i\beta}$ different from eq. (\ref{W1})
and $W^{\alpha\beta} \ne \delta^{\alpha\beta}$. 
Derivations of results in \S \ref{random} on estimator variance and 
the minimum variance
power spectrum estimator can be found in Appendix B.
We will ignore the integral constraint and assume $\bar
N^\alpha$ is known to high accuracy.

We need first of all the correlation matrix $\langle \hat
\delta_f^\alpha \hat \delta_f^\beta \rangle$.
We will do it in 2 steps, first let us derive $\langle \hat
\delta_f^\alpha \hat \delta_f^\beta \rangle_D$.
Rewriting $\hat \delta_f$ (eq. [\ref{hatdeltaf}]) as $(\hat N^\alpha -
\tilde N^\alpha)/\bar 
N^\alpha + (\tilde N^\alpha - \bar N^\alpha)/\bar
N^\alpha$ where $\tilde N^\alpha = \langle \hat N^\alpha \rangle_D$,
it can be shown that
\begin{eqnarray}
\langle \hat \delta_f^\alpha \hat \delta_f^\beta \rangle_D &=&
\delta_f^\alpha \delta_f^\beta + 
{1\over {\bar N_{\rm Q}^\alpha \bar N_{\rm Q}^\beta}} \langle (\hat
N^\alpha - \tilde N^\alpha) (\hat N^\beta - \tilde N^\beta) \rangle_D
\\ \nonumber
&=& \delta_f^\alpha \delta_f^\beta + {1\over {\bar N_{\rm Q}^\alpha
\bar N_{\rm Q}^\beta}} \sum_{\gamma 
i} W^{\alpha\gamma} W^{\beta\gamma} (W^{i\gamma})^2 (\langle \hat
N_{\rm Q}^{i\gamma} \rangle_D + V_{\rm B}^{i\gamma})
\label{pbrac}
\end{eqnarray}
where $\delta_f^\alpha$ is to be distinguished from $\hat
\delta_f^\alpha$ in that it has only cosmic or sample fluctuations
(eq. [\ref{Pxidef}]), $\hat N_{\rm Q}^{i\gamma}$ is a strictly
Poisson variable with an average given by eq. (\ref{ggNQ}), and
$V_{\rm B}^{i\gamma}$ is the variance contributions from the sky and
readout (eq. \ref{vRAW}). 

Taking the cosmic mean of the above, we obtain the correlation matrix
\begin{equation}
\langle \hat \delta_f^\alpha \hat \delta_f^\beta \rangle =
\xi (u^\alpha - u^\beta) + {1\over {\bar N_{\rm Q}^\alpha
\bar N_{\rm Q}^\beta}} \sum_{\gamma} W^{\alpha\gamma} W^{\beta\gamma}
\sum_i [\langle (W^{i\gamma})^2 g_{\rm ps}^{i\gamma} g_{\rm b}^\gamma
\tilde N_Q^\gamma \rangle + \langle (W^{i\gamma})^2 \rangle V_{\rm
B}^{i\gamma}]
\label{Cmatrix}
\end{equation}

The second term on the right is the shot-noise contribution which has
to be subtracted off. Using the estimator in eq. (\ref{hatP}), with
$w^{\alpha\beta} (k)$ given in eq. (\ref{wkbkuniform}), the
correct shot-noise subtraction is: 
\begin{equation}
b (k) = ({\cal L}/{\cal N}^2) \sum_{\alpha\beta\gamma}
W^{\alpha\gamma} W^{\beta\gamma} {\sum_i [\langle (W^{i\gamma})^2
g_{\rm ps}^{i\gamma} g_{\rm b}^\gamma 
\tilde N_{\rm Q}^\gamma \rangle + \langle (W^{i\gamma})^2 \rangle V_{\rm
B}^{i\gamma}] }
/ (\bar N_{\rm Q}^\alpha \bar N_{\rm Q}^\beta )
\label{bkgeneral}
\end{equation}
where we have made use of the assumption that $W^{\alpha\gamma}$ is
non-zero only for $\alpha$ and $\gamma$ on very small separations, and
that the $k$ of interest satisfies $k (u^\alpha - 
u^\gamma) \ll 1$ on such separations. 

Note that for weightings such as the one given in eq.
(\ref{W2}), $W^{i\gamma}$ depends on $\tilde N_{\rm Q}^\gamma$ which
makes an estimation of the shot-noise non-trivial. 
However, simplification results in two extreme limits: in the signal
dominated regime where $\tilde N_{\rm Q}^{j\beta} \gg V_{\rm
B}^{j\beta}$, $W^{i\gamma}$ reduces to uniform weighting as in
eq. (\ref{W1}); in the background dominated regime where
$\tilde N_{\rm Q}^{j\gamma} \ll V_{\rm
B}^{j\gamma}$, $W^{i\gamma}$ reduces to $(1/g_{\rm b}^\gamma) (g_{\rm
ps}^{i\gamma}/ V_{\rm B}^{i\gamma}) / (\sum_j (g_{\rm
ps}^{i\gamma})^2 /V_{\rm B}^{i\gamma})$ which is independent of $\tilde
N_{\rm Q}^\gamma$. In such cases, and for $W^{\alpha\gamma} =
\delta^{\alpha\gamma}$, the above $b (k)$ reduces to the one given in eq.
(\ref{bkuniform}). 

For $W^{\alpha\gamma} \ne \delta^{\alpha\gamma}$ (i.e. rebinning
has bee done; we continue to
assume $W^{i\gamma}$ is roughly independent of the signal),
we can write $b (k)$ as
\begin{equation}
b(k) \sim ({\cal L}/{\cal N}^2) \sum_{\alpha\beta\gamma}
{W^{\alpha\gamma} W^{\beta\gamma} \over 
{\bar N_{\rm Q}^\alpha \bar N_{\rm Q}^\beta}}
{\sum_i (W^{i\gamma})^2 [
g_{\rm ps}^{i\gamma} g_{\rm b}^\gamma 
\tilde N_{\rm Q}^\gamma \rangle + V_{\rm
B}^{i\gamma}]}
\label{bkgeneral2}
\end{equation}
where the term under summation of $i$ is simply the
variance array of the pre-rebinned data (eq. [\ref{varalpha}]),
and one can replace ${\bar N_{\rm Q}^\alpha} {\bar N_{\rm Q}^\beta}$
by $({\bar N_{\rm Q}^\alpha})^2$, since
$W^{\alpha\gamma} W^{\beta\gamma}$ is non-zero only if
$\alpha$ and $\beta$ are close together.
Eq. (\ref{bkapprox}) is therefore replaced by the following
if the data have been rebinned:
\begin{equation}
b(k) \sim {\Delta u \over {\cal N}} \sum_{\alpha\beta} 
{{\, \rm var \,} (\alpha,\beta)
\over {\bar N_Q^\alpha \bar N_Q^\beta}}
\label{bkapprox2}
\end{equation}
where ${\, \rm var \,} (\alpha,\beta)$ is the variance matrix
of the rebinned data.

To complete our derivation, we need to show that the choice of 
$w^{\alpha\beta} (k)$ given in eq. (\ref{wkbkuniform})
has the correct normalization such that the window function satisfies
$\int {dk' G(k,k')/{2\pi}} = 
1$ (eq. [\ref{Gwindow}]). Putting eq. (\ref{wkbkuniform}) into eq.
(\ref{hatP}), and using the correlation matrix given in eq.
(\ref{Cmatrix}) together with the relation between the two point
function and the power spectrum in eq. (\ref{Pxidef}), it is not hard
to see that $\langle \hat P_2 (k) \rangle$ satisfies eq.
(\ref{Gwindow}) with $G(k,k')$ given by 
$\sum_{\alpha\beta} w^{\alpha\beta} (k)
e^{ik'(u_\alpha- u_\beta)}$. Using $\int dk' e^{ik'(u_\alpha-
u_\beta)} = 2 \pi \delta^{\alpha\beta} {\cal N}/{\cal L}$ then
completes the derivation. One might want to explore more complicated
data windowing (e.g. \citenp{recipe92,hamilton97b}), but since in
practice uncertainties in the large scale power estimate, 
where the survey window matters most,
are likely dominated by the continuum, the simple
choice we have adopted is probably adequate.

\section*{Appendix B}
\label{app2}

We derive here the band-power variance given in eq. (\ref{Ek}).
The power spectrum estimator is given in eq. (\ref{hatP})
with the matrix $w^{\alpha\beta} (k)$ given by eq. (\ref{barw}).
We ignore here the uncertainty in the estimation of the
mean count $\bar N_{\rm Q}^\alpha$. 

The band-power covariance can be written compactly as
\begin{equation}
C (k_1, k_2) \equiv
\langle \Delta \hat P^\alpha (k_1) \Delta \hat P^\beta (k_2) \rangle
= \sum_{\alpha\beta\gamma\eta} w^{\alpha\beta} (k_1)
w^{\gamma\eta} (k_2) (\langle \hat \delta_f^\alpha \hat \delta_f^\beta 
\hat \delta_f^\gamma \hat \delta_f^\eta \rangle
 - \langle \hat \delta_f^\alpha 
\hat \delta_f^\beta \rangle \langle \hat \delta_f^\gamma
\hat \delta_f^\eta \rangle)
\label{C12}
\end{equation}
The band-power variance is simply the diagonal piece of
this matrix: $C(k, k)$.

We can work out $\langle \hat \delta_f^\alpha \hat \delta_f^\beta 
\hat \delta_f^\gamma \hat \delta_f^\eta \rangle
 - \langle \hat \delta_f^\alpha 
\hat \delta_f^\beta \rangle \langle \hat \delta_f^\gamma
\hat \delta_f^\eta \rangle$ using the same methodology
as used in Appendix A for $\langle \hat \delta_f^\alpha
\hat \delta_f^\beta \rangle$: rewrite $\hat \delta_f^\alpha$
as $(\hat N^\alpha -
\tilde N^\alpha)/\bar 
N^\alpha + (\tilde N^\alpha - \bar N^\alpha)/\bar
N^\alpha$ where $\tilde N^\alpha = \langle \hat N^\alpha \rangle_D$, and
as before, take the discrete-ensemble average $\langle\, \rangle_D$ before
taking the cosmic average $\langle \, \rangle$. 
The result is 
\begin{eqnarray}
\label{C12expand}
&& C (k_1, k_2) = 
\sum_{\alpha\beta\gamma\eta} w^{\eta\gamma} (k_1) 
w^{\alpha\beta} (k_2) \Biggl(
\\ \nonumber &&
\langle \delta_f^\alpha \delta_f^\beta \rangle
\langle \delta_f^\gamma \delta_f^\eta \rangle +
\langle \delta_f^\alpha \delta_f^\eta \rangle
\langle \delta_f^\beta \delta_f^\gamma \rangle +
\langle \delta_f^\alpha \delta_f^\gamma \rangle
\langle \delta_f^\beta \delta_f^\eta \rangle
+ \langle \delta_f^\alpha \delta_f^\beta \delta_f^\gamma \delta_f^\eta 
\rangle_c
%\sum_{\alpha} 
%\bar w^{\alpha\alpha} (k_1) \bar w^{\alpha\alpha} (k_2)
%{{\cal N}^3 \over {\cal L}^2} (
%{2 \over n_{k_1}} P(k_1)^2 \delta_{k_1 k_2} 
%+ \langle T \rangle_{k_1 k_2}) 
\\ \nonumber
&& + {1\over {\bar N_{\rm Q}^\eta \bar N_{\rm Q}^\gamma}}
\sum_{\sigma i}
W^{\eta\sigma} W^{\gamma\sigma} \langle (W^{i\sigma})^2
[g_{\rm ps}^{i\sigma} g_{\rm b}^\sigma \bar N_{\rm Q}^\sigma
(1 +\delta_f^\sigma) + V^{i\sigma}] \delta_f^\alpha \delta_f^\beta
\rangle \\ \nonumber &&
+ 1 {\, \, \rm other \, \, perm.:\,} (\alpha \leftrightarrow
\gamma, \beta \leftrightarrow \eta) \\ \nonumber
&& 
+ {1\over{\bar N_{\rm Q}^\beta 
\bar N_{\rm Q}^\gamma}} \sum_{\sigma i} W^{\beta\sigma} W^{\gamma \sigma}
\langle (W^{i\sigma})^2 (g_{\rm ps}^{i\sigma} g_{\rm b}^\sigma
\bar N_{\rm Q}^\sigma (1 + \delta_f^\sigma) + V_{\rm B}^{i\sigma})
 \delta_f^\alpha \delta_f^\eta 
\rangle \\ \nonumber
&&
+ 3 {\, \, \rm other \, \, perm.:} 
(\beta \leftrightarrow \alpha, \gamma \leftrightarrow \eta),
(\gamma \leftrightarrow \eta), (\beta \leftrightarrow \alpha)
\\ \nonumber
&& 
+ {1\over {\bar N_{\rm Q}^\beta \bar N_{\rm Q}^\gamma
\bar N_{\rm Q}^\eta}} \sum_{\sigma i} 
W^{\beta\sigma} W^{\eta\sigma} W^{\gamma\sigma}
\langle (W^{i\sigma})^3 [g_{\rm ps}^{i\sigma} g_{\rm b}^\sigma
\bar N_{\rm Q}^\sigma (1+ \delta_f^\sigma) 
+ \tilde N_{\rm S}^{i\sigma}]
\delta_f^\alpha \rangle \\ \nonumber &&
+ 3 {\, \, \rm other \, \, perm.:} 
(\alpha \leftrightarrow \beta), (\alpha \leftarrow \gamma),
(\alpha \leftrightarrow \eta)
\\ \nonumber
&&
+ {1\over {\bar N_{\rm Q}^\alpha 
\bar N_{\rm Q}^\beta \bar N_{\rm Q}^\gamma
\bar N_{\rm Q}^\eta}} \sum_{\sigma i\chi j}
W^{\alpha\sigma} W^{\beta\sigma} W^{\eta\chi} W^{\gamma\chi}
\langle (W^{i\sigma})^2 (W^{j\chi})^2
(g_{\rm ps}^{i\sigma} g_{\rm b}^\sigma
\bar N_{\rm Q}^\sigma (1 + \delta_f^\sigma) + V_{\rm B}^{i\sigma})
\\ \nonumber &&
(g_{\rm ps}^{j\chi} g_{\rm b}^\chi
\bar N_{\rm Q}^\chi (1 + \delta_f^\chi) + V_{\rm B}^{j\chi}) \\ \nonumber
&&
+ {1\over {\bar N_{\rm Q}^\alpha 
\bar N_{\rm Q}^\beta \bar N_{\rm Q}^\gamma
\bar N_{\rm Q}^\eta}} \sum_{\sigma i\chi j}
W^{\alpha\sigma} W^{\eta\sigma} W^{\beta\chi} W^{\gamma\chi}
\langle (W^{i\sigma})^2 (W^{j\chi})^2
(g_{\rm ps}^{i\sigma} g_{\rm b}^\sigma
\bar N_{\rm Q}^\sigma (1 + \delta_f^\sigma) + V_{\rm B}^{i\sigma})
\\ \nonumber &&
(g_{\rm ps}^{j\chi} g_{\rm b}^\chi
\bar N_{\rm Q}^\chi (1 + \delta_f^\chi) + V_{\rm B}^{j\chi}) 
+ 1 {\, \, \rm other \, \, perm.:}
(\alpha \leftrightarrow \beta)
\\ \nonumber
&&
+ {1\over {\bar N_{\rm Q}^\alpha 
\bar N_{\rm Q}^\beta \bar N_{\rm Q}^\gamma
\bar N_{\rm Q}^\eta}} \sum_{\sigma i}
W^{\alpha\sigma} W^{\beta\sigma} W^{\eta\sigma} W^{\gamma\sigma}
\langle (W^{i\sigma})^4 g_{\rm ps}^{i\sigma} g_{\rm b}^\sigma
\bar N_{\rm Q}^\sigma (1 + \delta_f^\sigma) \rangle \\ \nonumber
&& - \langle \hat \delta_f^\alpha \hat \delta_f^\beta \rangle
\langle \hat \delta_f^\gamma \hat \delta_f^\eta \rangle
\Biggr)
\end{eqnarray}

The first set of terms (second line) arise from the shot-noise-
free part of $\hat \delta_f^\alpha$, namely $(\tilde N^\alpha
- \bar N^\alpha) / \bar N^\alpha$. The next two sets of terms
(third + fourth and fifth + sixth lines) come from
combinations of $\hat \delta_f^\alpha$ involving products of
two shot-noise terms with two shot-noise-free terms.
The next set of terms (seventh + eighth lines) arises from
products of three shot-noise terms and a noise-free one.
The next set of terms (nineth to thirteenth lines)
comes from products of four shot-noise
terms. The last term corresponds to what has to be
subtracted off to compute the covariance.

To make further progress, we assume $W^{i\alpha}$ is independent
of $\delta_f^\alpha$, which is strictly correct for
$W^{i\alpha}$ given by eq. (\ref{W1}), but only roughly so
for eq. (\ref{W2}) (see discussion in Appendix A). 
Then, taking the small wavelength limit in the sense
that $k, \Delta k > 2\pi/{\cal L}$ ($\Delta k$ is the size
of a $k$-bin), and making use of the fact that 
$W^{\alpha\sigma} W^{\beta\sigma}$ is only non-zero at separations
$u^{\alpha} - u^{\beta}$ much less than $1/k$ where $k$ is the 
wavenumber of interest, we obtain
\begin{eqnarray}
\label{C12FKP}
&& C (k_1, k_2) = \sum_\sigma (\bar w^{\sigma\sigma})^2 {{\cal N}^3
\over {\cal L}^2} \Biggl(
\\ \nonumber
&& {2\over n_{k_1}} [P(k_1) + {{\cal L} \over {\cal N}}
\sum_{i} {1\over (\bar N_{\rm Q}^\sigma)^2}
\langle (W^{i\sigma})^2 (g_{\rm ps}^{i\sigma} g_{\rm b}^\sigma
\bar N_{\rm Q}^\sigma
+ V_{\rm B}^{i\sigma}) \rangle
\sum_{\beta\gamma} W^{\beta\sigma} W^{\gamma\sigma}]^2
\delta^{k_1 k_2} + {\langle T \rangle_{k_1 k_2} \over
{\cal L} }
\\ \nonumber
&& + 4 {1 \over {\cal N}} \langle B \rangle_{k_1 k_2}
\sum_{i} {1\over \bar N_{\rm Q}^\sigma}
\langle (W^{i\sigma})^2 g_{\rm ps}^{i\sigma} g_{\rm b}^\sigma \rangle
\sum_{\beta\gamma} W^{\beta\sigma} W^{\gamma\sigma}
\\ \nonumber
&& + 2 (P(k_1) + P(k_2)) {{\cal L} \over {\cal N}^2}
\sum_i {1\over (\bar N_{\rm Q}^\sigma)^2} \langle (W^{i\sigma})^3
g_{\rm ps}^{i\sigma} g_{\rm b}^\sigma \rangle 
\sum_{\beta\eta\gamma} W^{\beta\sigma} W^{\eta\sigma} W^{\gamma\sigma}
\\ \nonumber
&& + 2 \langle P \rangle_{k_1 k_2}
{{\cal L} \over {\cal N}^2}
\left[{1\over \bar N_{\rm Q}^\sigma} (W^{i\sigma})^2 g_{\rm ps}^{i\sigma}
g_{\rm b}^\sigma \sum_{\alpha\eta} W^{\alpha\sigma} W^{\eta\sigma} 
\right]^2
\\ \nonumber
&& + {{\cal L}^2 \over {\cal N}^3} {1\over (\bar N_{\rm Q}^\sigma)^4}
[(W^{i\sigma})^4 (g_{\rm ps}^{i\sigma} g_{\rm b}^\sigma
\bar N_{\rm Q}^\sigma + \tilde N_S^{i\sigma})]
\sum_{\alpha\beta\gamma\eta} W^{\alpha\sigma}
W^{\beta\sigma} W^{\eta\sigma} W^{\gamma\sigma} \Biggr)
\end{eqnarray}

A few comments are in order.
The terms in the third + fourth + nineth + tenth lines
of eq. (\ref{C12expand}) are canceled by the last term of
eq. (\ref{C12expand}). The terms in the third + fourth
lines of eq. (\ref{C12expand}) contain contributions proportional to
$B(k,-k,0)$ which vanish.
There is also a term from the nineth line of eq. (\ref{C12expand}) that is
proportional to $P(0)$ which vanishes also.
The shot-noise terms in the second line of
eq. (\ref{C12FKP}) come respectively from
the fifth + sixth + nineth + tenth lines of eq. (\ref{C12expand}).
The rest of the terms in eq. (\ref{C12FKP}) basically
follow the order they are presented in eq. (\ref{C12expand}).

Lastly, setting $W^{\alpha\beta} = \delta^{\alpha\beta}$ and
$k_1 = k_2$ then
recovers eq. (\ref{Ek}). 

%\bibliographystyle{astro}
%\bibliography{numpdsnew}

\newpage

%NOnoise.plot
%shows simulated spectrum from 
%/d/sculptor/lhui_sculptor/PDS/NewErr/MoreData/Long0.037_NO.noise.gap.cont
%the power spectrum is from
%/d/honus/lhui/Run/MoreRun/Long0.037_NO.noise.gap.cont/N1_Nkk30_Ispds0tr0
%seems O.K.

%highKeckNM.plot
%0.01619820856      <sigma>
%0.6937941313       <signal>
%0.0005450953031    (<sigma>/<signal>)**2
%0.0007071972359    NN = (Nqbar*fb + Nsky*fb)/Nqbar**2/fb/fb
%0.0498046875       (dr)
%                   i.e. shot-noise term is dr * NN, should
%                   be compared with P(k), can compare dr * NN
%                   with bias.dat in the analysis directory
%spectrum from 
%/d/sculptor/lhui_sculptor/PDS/NewErr/MoreData/Long0.037_simerr5_cr_Nf1_cont2NM
%signal = third column in spec.dat / second column in blaze.dat
%input continuum is analytic. 0.64 * continuum gives the true mean.
%noise = (third column in spec.dat + second column in sky.dat)**0.5/
%        (second column in blaze.dat)
% the data here have small cosmic-ray gaps (3% of all pixels).
%
% the estimated mean's are from:
%/d/lyra/lhui/Run/MoreRun/Long0.037_simerr5_cr_Nf1_cont2NM/N1_Nkk400_Ispds0
%/d/lyra/lhui/Run/MoreRun/Long0.037_simerr5_cr_Nf1_cont2NM/N4_Nkk400_Ispds0
%/d/lyra/lhui/Run/MoreRun/Long0.037_simerr5_cr_Nf1_cont2NM/
% N4_Nkk400_Ispds0Istrend1
%The measurements of the power spectrum from the above directories
%are then compared to
%/d/lyra/lhui/Run/MoreRun/Long0.037_NO.noise.gap.cont/N1_Nkk400_Ispds0
%
%*Should rerun the noisy runs with integral-constraint bias taken out
%and see what happens. but don't expect a big change here because
%the error is large anyway.
%*Actually, I have decided that the integral-constraint bias
% is not big (~ 1-2 %), and so will not worry about it;
% see - /d/lyra/lhui/LA_compute/Analyze/Test3_31_99/Runs
%       /NO.noise.gap.cont_MultiRealiz/All/N1_Nkk20New_Ispds1_J0.049
%       /pdscheckout.plot

%highKeckNM.C.plot
%spectrum from 
%/d/sculptor/lhui_sculptor/PDS/NewErr/MoreData/Long0.037_simerr6_cr_Nf1_cont3NM
%continuum is from cont3.dat
%
%estimated mean's from
%/d/lyra/lhui/Run/MoreRun/Long0.037_simerr6_cr_Nf1_cont3NM
% /N1_Nkk20New_Ispds1Istr1
%/d/lyra/lhui/Run/MoreRun/Long0.037_simerr6_cr_Nf1_cont3NM
% /N4_Nkk20New_Ispds1Istr1
%
% DP/P computed using as true power spectrum
%/d/lyra/lhui/Run/MoreRun/Long0.037_NO.noise.gap.cont
% /N1_Nkk20New_Ispds1Istrend1, instead of /N1_Nkk20New_Ispds1.

%combining the old highKeckNM.C_S.plot and highKeckNM.C_S2.plot
%highKeckNM.C_S.plot

%combined the old highKeckNM.C_S.z4.plot and highKeckNM.C_S2.z4.plot
%highKeckNM.C_S.z4.plot

%combined the old LickNM_cont3L_S.plot and LickNM_cont3L_S2.plot
%LickNM_cont3L_S.plot

%LickNM_cont3L.plot

%see pdsScottCont3.plot (the old version used pdsScottCont.plot)
%and /d/lyra/lhui/LA-compute/Analyze/Test3-31-99/guide, look for
%Runs/Cont
%pdsScottCont2.plot (the old version used pdsScottCont.plot)
%QSO 1157+3143  
%see Workspace/lhui/Fermi/honus/lhui/Theory/Da/Corr/9_4/Another

%combKeck.plot

%combLick.plot

%this combined the old highKeckM.C.plot and highKeckMgap.C.plot
%highKeckM.C.plot
%the metal lines are taken originally from
%/work/lhui/Workspace/Fermi/honus/lhui/Theory/Da/Corr/7_31

%this combined the old highKeckM.C.plot and highKeckMgap.C.plot
%LickM_cont3L.plot

%based on OLD highKeckNM.C.FKP.plot
%highKeckNM.C.FKP2.plot

%highKeckNM.C_S2.plot
%\begin{figure}[htb]
%\centerline{\psfig{figure=highKeckNM.C_S2.ps,height=3.0in}}
%\caption{highKeckNM.$C_S2$.plot} 
%\label{spec}
%\end{figure}

%highKeckNM.C_S2.z4.plot
%\begin{figure}[htb]
%\centerline{\psfig{figure=highKeckNM.C_S2.z4.ps,height=3.0in}}
%\caption{highKeckNM.$C_S2$.z4.plot} 
%\label{spec}
%\end{figure}

%LickNM.plot
%\begin{figure}[htb]
%\centerline{\psfig{figure=LickNM.ps,height=3.0in}}
%\caption{LickNM.plot, with and without shot-noise correction.} 
%\label{spec}
%\end{figure}

%LickNM_cont3L_S2.plot
%\begin{figure}[htb]
%\centerline{\psfig{figure=LickNM_cont3L_S2.ps,height=3.0in}}
%\caption{LickN$M_c$ont3$L_S$2.plot, un-normalized power.} 
%\label{spec}
%\end{figure}

%highKeckMgap.C.plot
%\begin{figure}[htb]
%\centerline{\psfig{figure=highKeckMgap.C.ps,height=3.0in}}
%\caption{highKeckMgap.C.plot, dotted: with metal;
%long-dashed: strong metal taken out with gaps} 
%\label{spec}
%\end{figure}

%Var.plot
%\begin{figure}[htb]
%\centerline{\psfig{figure=Var.ps,height=3.0in}}
%\caption{Var.plot} 
%\label{spec}
%\end{figure}

%use demonstrate.plot

%%use combKeckB.plot
%%based on runs in 
%%/work/lhui/Workspace/Fermi/lyra/lhui/LA_compute/Analyze/Test3_31_99/
%%Runs/Cont/FlatCont --> see jobtotal
%\begin{figure}[htb]
%\centerline{\psfig{figure=combKeckB.ps,height=3.0in}}
%\caption{This is a figure purely for my own reference. It will
%not be included in the final paper. It shows an analysis of
%the spectrum without any noise or continuum added (i.e. a
%flat continuum, as shown in the second panel) -- but
%I ``mistakenly'' perform trend-removal as usual with 
%an assumed basis consisting of up to third order polynomials.
%Such a procedure would recover the true continuum no problem,
%when one is presented with the actual (flat) continuum
%as shown by a dotted line in the second panel (indistinguishable
%from the solid line). But when
%applied to the forest, the procedure removes power on large
%scales, as shown by the solid line in the top panel.
%Power correction by subtraction of the continuum power alone
%will not do, since the continuum power is basically zero
%(dotted line which overlaps with solid line). However,
%correction using $\epsilon_G$ (dashed line) restores the
%large scale power.}
%\label{combKeckB}
%\end{figure}

\end{document}